\documentclass[authoryear,5p]{elsarticle}

\usepackage{natbib}

\usepackage{epsfig}

\usepackage{amssymb}

\begin{document}

\begin{frontmatter}



\title{Measurement of neutron star parameters: a review of methods for low-mass X-ray binaries}


\author{Sudip Bhattacharyya}
\address{Department of Astronomy and Astrophysics, Tata Institute of Fundamental Research, Mumbai 400005, India
\ead{sudip@tifr.res.in}}

\begin{abstract}
Measurement of at least three independent parameters, 
for example, mass, radius and spin frequency,
of a neutron star is probably the only way to understand the nature of its
supranuclear core matter. Such a measurement is extremely difficult because 
of various systematic uncertainties. The lack of knowledge of several system
parameter values gives rise to such systematics. Low mass X-ray binaries, which
contain neutron stars, provide a number of methods to constrain the stellar 
parameters. Joint application of these methods has a great potential
to significantly reduce the systematic uncertainties, and hence to
measure three independent neutron star parameters accurately. Here we review
the methods based on (1) thermonuclear X-ray bursts; (2) accretion-powered 
millisecond-period pulsations; (3) kilohertz quasi-periodic oscillations; 
(4) broad relativistic iron lines; (5) quiescent emissions; and
(6) binary orbital motions.
\end{abstract}


\end{frontmatter}

\parindent=0.5 cm

\noindent
{\large\bf Contents}

\noindent
{\bf 1 Introduction}\\
{\bf 2 Why This Review and Why Now?}\\
{\bf 3 Methods}\\
\hspace*{0.3cm} 3.1 Thermonuclear X-ray Bursts\\
\hspace*{0.6cm} 3.1.1 What is a Thermonuclear X-ray Burst\\
\hspace*{0.6cm} 3.1.2 Why are Thermonuclear X-ray Bursts Useful?\\
\hspace*{0.6cm} 3.1.3 Continuum Spectrum Method\\
\hspace*{0.6cm} 3.1.4 Spectral Line Method\\
\hspace*{0.6cm} 3.1.5 Photospheric Radius Expansion Burst Method\\
\hspace*{0.6cm} 3.1.6 Burst Oscillation Method\\
\hspace*{0.6cm} 3.1.7 Millihertz Quasi-periodic Oscillation Method\\
\hspace*{0.6cm} 3.1.8 Joint Application of the Methods based on\\
\hspace*{1.4cm} Thermonuclear X-ray Bursts\\
\hspace*{0.3cm} 3.2 Accretion-powered Millisecond Period Pulsations\\
\hspace*{0.6cm} 3.2.1 What is an Accretion-powered Millisecond\\
\hspace*{1.4cm} Pulsar?\\
\hspace*{0.6cm} 3.2.2 Accretion-powered Millisecond Pulsation Method\\
\hspace*{0.3cm} 3.3 Kilohertz Quasi-periodic Oscillations\\
\hspace*{0.6cm} 3.3.1 What are Kilohertz Quasi-periodic Oscillations?\\
\hspace*{0.6cm} 3.3.2 Kilohertz Quasi-periodic Oscillation Method\\
\hspace*{0.3cm} 3.4 Broad Relativistic Iron Lines\\
\hspace*{0.6cm} 3.4.1 What is a Broad Relativistic Iron Line\\
\hspace*{0.6cm} 3.4.2 Broad Relativistic Iron Line Method\\
\hspace*{0.3cm} 3.5 Quiescent Emissions\\
\hspace*{0.6cm} 3.5.1 What is Quiescent Emission?\\
\hspace*{0.6cm} 3.5.2 Quiescent Emission Method\\
\hspace*{0.3cm} 3.6 Mass Measurement: Binary Orbital Motion Method\\
{\bf 4 Why are Low-mass X-ray Binaries Useful?}\\
{\bf 5 Summary and Future Prospects}

\section{Introduction}\label{Introduction}

The nature of matter at ultra-high density, i.e., $5-10$ times higher than the
nuclear density, but at a relatively low temperature (e.g., $\sim 10^8$~K) cannot
be probed by heavy-nuclei collision experiments or with observations of the
early universe (\cite{OzelPsaltis2009, vanKerkwijk2004, 
Blaschkeetal2008, LattimerPrakash2007} and references therein).
This matter exists in the neutron star cores, and hence these stars provide
natural laboratories to study the supranuclear degenerate matter. The study of
this dense matter is not only important from the point of view of the 
particle physics, but also can be useful to understand the core collapse of
massive stars, the supernova phenomenon, the existence and properties
of neutron stars, etc. (e.g., \cite{vanKerkwijk2004}).

\begin{figure}[h]
\begin{center}
\hspace*{-0.8cm}
\includegraphics*[width=10cm,angle=0]{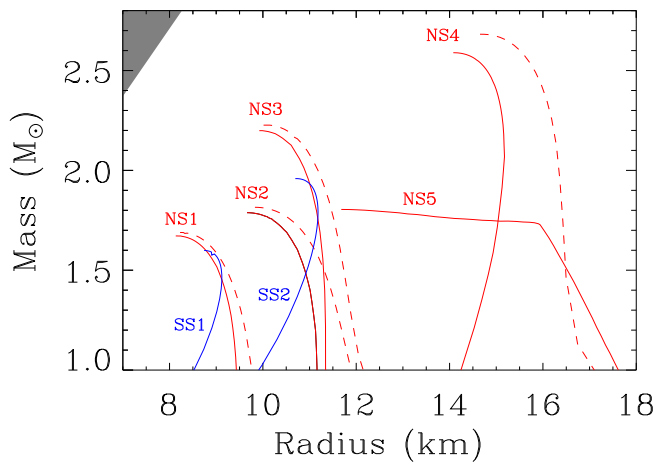}
\end{center}
\caption{This figure shows the mass-radius ($M-R$) space of neutron stars and the curves
corresponding to a few representative EoS models (\S~\ref{Introduction}). 
The solid curves are for nonspinning stars, while
the dashed curves (corresponding to some of the solid curves) are for 
a stellar spin frequency of 600 Hz. The red curves are for the usual neutron star
EoS models: NS1 (A18 model of \cite{Akmaletal1998}); NS2 \citep{Baldoetal1997}; 
NS3 (A18+$\delta v$+UIX model of \cite{Akmaletal1998}); NS4 \citep{Sahuetal1993}; 
and NS5 \citep{PandharipandeSmith1974}; while the blue 
curves are for the strange matter EoS models: SS1 (bag constant $= 90$ MeV fm$^{-3}$; 
\cite{FarhiJaffe1984}) and SS2 (bag constant $= 60$ MeV fm$^{-3}$; \cite{FarhiJaffe1984}; 
see also \S~\ref{Introduction}). The grey region indicates the $M-R$ space of
black holes. 
Here we have shown the $M > 1~M_{\odot}$ region, because many neutron stars
in double neutron star binaries
have the mass around $1.35~M_{\odot}$ \citep{ThorsettChakrabarty1999}, 
and therefore the accreting neutron stars in low-mass X-ray binaries
should have even larger mass values (see \S~\ref{WhyareLow-massX-rayBinariesuseful}).
This figure suggests that the measurement of mass, radius and
spin frequency of the same neutron star is required to constrain the EoS models.
\label{eosfirst}}
\end{figure}

But how to observe and understand the deep interior of a neutron star?
Based on the assumed microscopic properties of the core matter, many
equation of state (EoS) models have been proposed in the literature.
These models connect the pressure ($p$) with the total mass-energy 
density ($\epsilon$) in a degenerate condition.
Therefore, for a nonspinning neutron star and for a given EoS model, the stellar
global structure can be calculated by solving the Tolman-Oppenheimer-Volkoff
(TOV) equation:
\begin{eqnarray}
\frac{{\rm d}p(r)}{{\rm d}r} = -\frac{G}{c^2}\frac{[p(r)+\epsilon(r)][m(r)+4\pi r^3p(r)/c^2]}{r(r-2Gm(r)/c^2)},
\label{TOV}
\end{eqnarray}
where, $m(r) = 4\pi \int_0^r {\rm d}r^{\prime}r^{\prime 2}\epsilon(r^{\prime})$
is the gravitational mass (hereafter, mass) inside a sphere of radius $r$.
For a rapidly spinning neutron star, \citet{Cooketal1994} and other authors have
suggested ways to compute the stellar structure (see the dashed curves
of Figure~\ref{eosfirst}; see also, 
\cite{Dattaetal1998, Thampanetal1999, Bhattacharyyaetal2000, 
BhattacharyyaBhattacharyaThampan2001, BhattacharyyaMisraThampan2001}). 
The Figure~\ref{eosfirst} shows that for a given EoS model and a known stellar
spin frequency the stable structures of neutron stars trace a single 
mass($M$)-radius($R$) curve. However, such curves for different EoS models
can intersect each other (Figure~\ref{eosfirst}). Therefore, it is required to
accurately measure at least three
independent global parameters of the {\it same} neutron star 
in order to constrain the EoS models,
and hence to understand the supranuclear core matter.
This is also true for the EoS models of strange matter (made of $u$, $d$ and $s$ quarks;
\cite{Witten1984, FarhiJaffe1984,
Bombaci1997, Bombacietal2000,
BhattacharyyaThampanBombaci2001, Bagchietal2006}).
However, some neutron star EoS models can be ruled out only by mass
measurement. Here is how. 
A stable neutron star's mass cannot be greater than a maximum value ($M_{\rm max}$;
\cite{LattimerPrakash2007}) for a given EoS model and spin frequency.
Therefore, a large measured
mass value can rule out some ``softer" EoS models (for example, ``NS1" in
Figure~\ref{eosfirst}), that have lower $M_{\rm max}$ than the ``harder" models 
(for example, ``NS4" in Figure~\ref{eosfirst}; see also \cite{LattimerPrakash2007}).

The above discussion give the motivation for an accurate measurement of the neutron 
star parameters. In this review, we discuss the measurement methods 
relevant for neutron stars in low-mass X-ray binary (LMXB) systems 
\citep{BhattacharyaHeuvel1991}.
An LMXB is an old binary stellar system (typical age
$\sim 10^9$ years) with a low-mass companion star
($\le 1M_{\odot}$) and a compact star rotating around each other.
The companion star may be a main sequence star, an evolved star or a white dwarf,
while the compact star may be a neutron star or a black hole. In such
a system, the companion star fills its Roche lobe, i.e., the critical equipotential
surface (see \cite{BhattacharyaHeuvel1991}), and matter from it flows 
towards the compact star (neutron star in our case). 
This matter cannot move radially towards the neutron star,
as its initial angular momentum is large. Therefore, the accreted
matter slowly gets rid of the angular momentum \citep{ShakuraSunyaev1973},
follows spiral paths, and hence forms an
accretion disk. In neutron star LMXBs, the accreted matter
eventually hits the stellar surface, and generates electromagnetic radiation.
The accretion disk also emits such radiation by viscous dissipation, and
both these emissions are powered by the gravitational potential energy
release. The inner part of the accretion disk, and the neutron star
surface emit X-rays, as their typical temperature is $\sim 10^7$ K.
These X-rays are reprocessed by the outer part of the accretion disk,
which emits in optical wavelengths. Moreover, the companion star
may be irradiated by these X-rays.
The neutron stars in LMXBs are expected to be spun up, due to the
angular momentum transferred via accretion. This is consistent with the
observed stellar spin frequencies (typically $\sim 300-600$ Hz; see 
\S~\ref{BurstOscillationMethod}; \S~\ref{Accretion-poweredMillisecondPulsationMethod}).
The magnetic fields ($\approx 10^7-10^9$ G) of these neutron stars are typically 
much lower than that of the isolated neutron stars and the
stars in high-mass X-ray binary (HMXB) systems \citep{BhattacharyaHeuvel1991,
IbragimovPoutanen2009, PsaltisChakrabarty1999}.
Several groups have worked on this problem of
stellar magnetic field evolution \citep{Srinivasanetal1990,
JahanMiriBhattacharya1995, Konaretal1995, ChoudhuriKonar2002, 
KonarChoudhuri2004}. Note that the low magnetic
field allows the accretion disk to extend up to very close to the neutron star.
This, in combination with the fact that X-rays can be detected both from the
star and the inner part of the disk for many LMXBs, provides an excellent opportunity
to measure the neutron star parameters.
Moreover, the signatures of binary orbital motion detected from the
optical observation of the accretion disk and the companion star
can be useful to measure the neutron star mass.

LMXBs are observed mostly in the Galactic plane and bulge, and in the globular clusters.
Some of them have also been discovered from the nearby galaxies. According to the
catalogue of \citet{Liu2007}, 187 LMXBs (containing either a
neutron star or a black hole)
were known from our galaxy, ``Large Magellanic Cloud" and 
``Small Magellanic Cloud" up to that time. 
As indicated from thermonuclear bursts, mass function, etc., most of these
LMXBs contain a neutron star, rather than a black hole.
LMXBs emit primarily in X-ray wavelengths, and the optical to X-ray
luminosity ratio is normally less than 0.1 for them (see \cite{BhattacharyaHeuvel1991}).
The central X-ray emission from these sources cannot be spatially resolved
using the current X-ray imaging instruments, because the distances
of even our Galactic LMXBs are very large (typically $> 1$ kpc).
Therefore, one needs to rely on the
spectral and timing properties in order to study these sources in X-rays.

X-rays from space cannot reach the surface of the Earth. Therefore, 
the X-ray detectors and telescopes are sent above the atmosphere, typically
by satellites.
The current primary X-ray space missions are (1) NASA's {\it Rossi X-ray Timing Explorer}
({\it RXTE}); (2) ESA's {\it XMM-Newton}; (3) NASA's {\it Chandra}; and
(4) JAXA's {\it Suzaku}. They primarily operate in the $\approx 1-10$ keV range
(i.e., soft X-rays), and contain a variety of instruments.
The primary instrument of {\it RXTE} is a set of five
proportional counters (called ``Proportional Counter Array" or PCA) with large
collecting area and very good time resolution ($\approx 1$ microsecond). Therefore,
PCA is an ideal instrument to study the fast timing phenomena, such as
burst oscillations (\S~\ref{BurstOscillationMethod}), accretion-powered pulsations 
(\S~\ref{Accretion-poweredMillisecondPeriodPulsations}),
kilohertz quasi-periodic oscillations (\S~\ref{KilohertzQuasi-periodicOscillations}), 
and also the continuum energy spectrum (e.g., \S~\ref{ContinuumSpectrumMethod}).
PCA is also the best instrument to study most of the properties of
thermonuclear bursts (see \S~\ref{ThermonuclearX-rayBursts}). 
However, it is not suitable to study the spectral lines because of its
poor energy resolution. This instrument operates in $2-60$ KeV band, 
although the effective collecting area becomes small above $\approx 20$ keV.
{\it XMM-Newton} satellite
contains X-ray telescopes, charge-coupled devices and high resolution
``Reflection Grating Spectrometers (RGS)". Therefore, this satellite,
which operates in $\approx 0.2-12$ keV range, is ideal for spectral analysis,
including narrow spectral lines.
The {\it Chandra} observatory contains an X-ray telescope, charge-coupled devices,
high resolution camera and transmission gratings. Its angular resolution ($\sim$
subarcsecond) is the best among all the past and current X-ray instruments, and
it is also ideal to detect and study narrow spectral lines (in $\approx 0.1-10$ 
keV range). {\it XMM-Newton} and {\it Chandra} are also useful to study
the quiescent emission of neutron star LMXBs (\S~\ref{QuiescentEmissions}).
The {\it Suzaku} satellite also has X-ray telescopes and 
charge-coupled devices. Since it contains an additional ``Hard X-ray Detector" 
(HXD), it can study the broadband energy spectrum.
Therefore, {\it Suzaku} is very useful to observe the broad relativistic iron
emission lines and the associated ``disk reflection'' spectra
(see \S~\ref{BroadRelativisticIronLines}).

Here is the plan of this review. In \S~\ref{WhyReview}, we describe the aim and
the timeliness of this review. In \S~\ref{Methods}, we briefly discuss a few
observed properties of neutron star LMXBs, how these properties can be used 
to measure neutron star parameters, the relevant systematics and the plausible
ways to reduce these systematic uncertainties. \S~\ref{WhyareLow-massX-rayBinariesuseful}
mentions why LMXBs are useful to measure the neutron star parameters,
and \S~\ref{SummaryandFutureProspects} includes a summary of this review,
as well as a brief discussion on future prospects.

\section{Why This Review and Why Now?}\label{WhyReview}

This review aims to describe the major neutron-star-parameter-measurement 
methods for LMXB systems briefly, and to be useful as a quick reference for
astrophysicists, nuclear physicists, particle physicists, and the
scientists working in fluid mechanics, general theory of relativity,
and other relevant fields. This is probably the first time that all the methods,
including a recently established method based on
broad relativistic spectral iron emission lines 
(see \S~\ref{BroadRelativisticIronLineMethod}),
for LMXB systems have exclusively been compiled together.
Although, this review does not include the techniques for all
types of neutron star systems,
the discussed methods form a complete set for the
purpose of constraining EoS models. This is because at least
three independent parameters of the {\it same} neutron star must
be measured to constrain these models (see \S~\ref{Introduction}).

We believe that this review is timely, because of several new 
discoveries and advancement in the field of neutron star LMXBs made in
the past few years. To mention a few, (1) \citet{Chakrabartyetal2003}
established that the burst oscillations do give the neutron star
spin frequency (see \S~\ref{BurstOscillationMethod}); 
(2) the sample size of accretion-powered millisecond-period pulsars
has increased from 1 to 12 (\S~\ref{WhatisanAccretion-poweredAccretion-poweredPulsar});
(3) \citet{Cottametal2002} reported the only plausible observation of the 
neutron star surface atomic spectral lines (see \S~\ref{SpectralLineMethod});
(4) \citet{BhattacharyyaStrohmayer2007b} established the inner accretion disk 
(or relativistic) origin of the broad iron line from a neutron star LMXB for 
the first time (see \S~\ref{WhatisaBroadRelativisticIronLine}); and
(5) \citet{SteeghsCasares2002} detected the companion star of an 
LMXB based on Bowen blend emission lines, and constrained the neutron
star mass using these lines for the first time 
(see \S~\ref{MassMeasurementUsingBinaryOrbitalMotion}).

\section{Methods}\label{Methods}

X-ray emission of an LMXB from a region close to the neutron star
surface contains information about the stellar parameters.
This information can be extracted from the observed spectral 
and timing properties of these X-rays. Here we discuss the
properties of the following five X-ray phenomena,
and the methods for neutron star parameter measurement using them:
(1) thermonuclear X-ray bursts; (2) accretion-powered millisecond-period pulsations;
(3) kilohertz quasi-periodic oscillations; (4) broad relativistic iron lines;
and (5) quiescent emissions.
In addition, we discuss a mass measurement technique based on
the orbital motions of the binary components.

\subsection{Thermonuclear X-ray Bursts}\label{ThermonuclearX-rayBursts}

\subsubsection{What is a Thermonuclear X-ray Burst?}\label{WhatisaThermonuclearX-rayBurst}

\begin{figure}[h]
\begin{center}
\includegraphics*[width=8cm,angle=0]{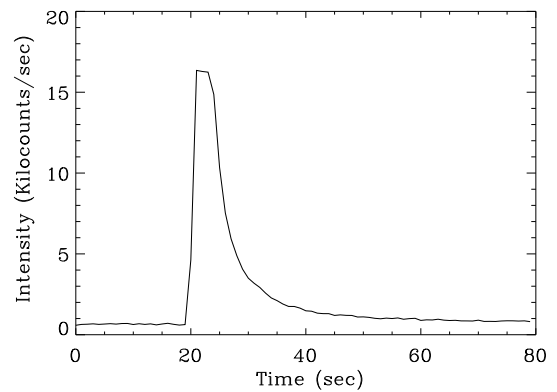}
\end{center}
\caption{
This figure shows the X-ray intensity profile of 
a typical thermonuclear X-ray burst observed from an LMXB.
The burst light curve is for the entire {\it RXTE} PCA band, and with 
1 sec binsize. After the onset of the burst, the X-ray intensity
increases sharply by more than an order of magnitude compared to the pre-burst
intensity (powered by the gravitational potential energy), and then
decays relatively slowly as the neutron star surface cools down
(\S~\ref{WhatisaThermonuclearX-rayBurst}).
\label{burst2}}
\end{figure}

Eruptions in X-rays are observed from many neutron star 
LMXB systems every few hours to days \citep{StrohmayerBildsten2006,
Galloway2008}. These are called type-I X-ray bursts.
The observed X-ray intensity sharply increases typically by a factor of
$\sim 10$ in $\approx 0.5-5$ seconds, and then
decreases relatively slowly in $\approx 10-100$ seconds during such a burst
(\cite{Galloway2008}; see also Figure~\ref{burst2}). The typical energy emitted in a 
few seconds is $\sim 10^{39}$ ergs 
\citep{StrohmayerBildsten2006}. These bursts were first discovered 
in 1970's \citep{Grindlay1976, Belian1976}.
Soon after this discovery it was realized that they
originate from intermittent unstable nuclear burning of accreted matter 
on the neutron star surfaces \citep{Joss1977, LambLamb1978,
StrohmayerBildsten2006}.
The neutron star surface origin was supported by the observational
fact that the burst emission area matched with the expected surface area
of a neutron star \citep{Swank1977, Hoffmanetal1977a}.
Note that the emission area
can be estimated from the fitting of the energy spectra of the bursts.
The unstable burning is required because, while the
nuclear energy released per nucleon by fusion is only a few MeV,
the gravitational energy released by a nucleon falling on the neutron star 
surface from a large distance is $GMm_{\rm nucleon}/R$ ($M$ and $R$ 
are neutron star mass and radius respectively), which is typically 
$\sim 200$ MeV. Therefore, if the nuclear burning were stable, the 
nuclear energy would be entirely lost in the observed gravitational 
energy, and the burst could not be powered by the nuclear energy 
release. This nuclear burning instability is driven by a nuclear energy 
generation rate which is more temperature sensitive than the radiative cooling.
Such a ``thin shell" instability was first theoretically
discovered by \citet{Schwarzschild1965} for the asymptotic giant
branch phase of stellar evolution, and was shown to 
occur in the thin shell of hydrogen and helium
on neutron star surfaces \citep{Hansen1975}.
But what shows that the nuclear burning causes the bursts?
A primary evidence of it comes from the following observational fact.
For a data set from a given source,
the ratio of the total energy content in all the bursts to that in the non-burst
emission roughly tallies with the ratio of the expected nuclear energy 
release per nucleon to the expected gravitational energy release per nucleon.
This strongly suggests that the accreted matter accumulates on the neutron
star surface for a long time, and then produces a burst via a quick 
thermonuclear burning.
Therefore these bursts are known as ``thermonuclear bursts"
(see the reviews \cite{Lewin1993, Lewin1995} and 
\citet{StrohmayerBildsten2006} for detailed descriptions of these bursts).

For many neutron star LMXBs, the accreted material consists of mostly 
hydrogen, some helium and a small amount of heavier elements, although
for ultra-compact X-ray binaries (UCXBs) the donor companion 
is a hydrogen deficient and heavier element rich dwarf star 
(\cite{zand2008} and references therein).
The accreted matter, which accumulates on the neutron star surface, undergoes
hydrostatic compression as more and more material piles on,
and the ignition column density
and temperature are reached in a few hours to days \citep{SchatzRehm2006}.
The burning typically happens at a depth of $\approx 10$ m at a column density
of $\sim 10^8$ gm cm$^{-2}$. The compression rate
and the ignition depends on the accretion rate, which sets four distinct
theoretically expected regimes of burning \citep{Bildsten2000}.\\
{\it Regime 1}: The temperature of the accumulated material
exceeds $10^7$ K in most cases, and hence the hydrogen
burns via the CNO (carbon-nitrogen-oxygen) 
cycle instead of the pp (proton-proton) cycle.
This causes mixed hydrogen and helium bursts triggered by the unstable
hydrogen ignition, which happens for the accretion rate per unit neutron star
surface area \.{m} $ < 900$ gm cm$^{-2}$ sec$^{-1}$ ($Z_{\rm CNO}/0.01$)$^{1/2}$
\citep{Bildsten2000}. Here $Z_{\rm CNO}$ is the mass fraction of CNO.\\
{\it Regime 2}: the proton capture timescale becomes shorter than the 
subsequent $\beta$ decay lifetimes at higher temperatures 
($T > 8\times10^7$ K). Hence, the hydrogen
burns via the ``hot" CNO cycle of \citet{Fowler1965}:
$$^{12}{\rm C}(p,\gamma)^{13}{\rm N}(p,\gamma)^{14}{\rm O}(\beta^{+})^{14}{\rm N}(p,\gamma)^{15}{\rm O}(\beta^{+})^{15}{\rm N}(p,\alpha)^{12}{\rm C}.$$
In this case, the hydrogen burning happens
in a thermally stable manner (i.e., without triggering
a thermonuclear burst) simultaneously with the accumulation of matter.
This allows a helium layer to build up below the hydrogen layer. For the
\.{m} range $900$ gm cm$^{-2}$ sec$^{-1}$ ($Z_{\rm CNO}/0.01$)$^{1/2}
< $ \.{m} $< 2\times10^3$ gm cm$^{-2}$ sec$^{-1}$ ($Z_{\rm CNO}/0.01$)$^{13/18}$,
the hydrogen entirely burns before this helium is ignited
\citep{Bildsten2000, StrohmayerBildsten2006}. So when
the helium ignition happens, a short ($\sim 10$ sec) but very intense burst occurs by the
unstable triple-alpha reaction of the pure helium ($3\alpha$ $\rightarrow$ $^{12}{\rm C}$).
Such a burst is called a ``helium burst". Many of these bursts are so intense
that the local X-ray luminosity in the neutron star atmosphere may exceed the
Eddington limit (for which the radiative pressure force balances the gravitational
force), and the photospheric layers may be lifted off the neutron star surface.
Such bursts are called photospheric radius expansion bursts 
(\cite{StrohmayerBildsten2006} and references therein).\\
{\it Regime 3}: hydrogen burns via the ``hot" CNO cycle in a stable manner 
for \.{m} $> 2\times10^3$ gm cm$^{-2}$ sec$^{-1}$
($Z_{\rm CNO}/0.01$)$^{13/18}$ (as for the regime 2). 
But in this case, enough amount of unburnt hydrogen remains
present at the time of helium ignition. This is because at this higher
accretion rate, the helium ignition conditions are satisfied much sooner.
Therefore in this regime, mixed hydrogen and helium bursts are triggered
by the helium ignition. During such a burst, the thermal instability
can produce elements beyond the iron group \citep{Hanawa1983,
HanawaFujimoto1984, Koike1999, 
Schatz2001} via the rp (rapid-proton) process of \citet{WallaceWoosley1981}. 
This is because the triggering helium flash makes the temperature
high enough to allow the ``breakout reaction" $^{15}{\rm O}(\alpha,\gamma)^{19}{\rm 
Ne}$ \citep{WallaceWoosley1981, Schatz1999, Fisker2006} from hot CNO cycle
to proceed faster than the $\beta$ decays. $^{19}{\rm Ne}$ can return to 
the hot CNO cycle via the chain of reactions \citep{Cooper2006}:
$$^{19}{\rm Ne}(\beta^{+}\nu)^{19}{\rm F}(p,\alpha)^{16}{\rm O}(p,\gamma)^{17}{\rm
F}(p,\gamma)^{18}{\rm Ne}(\beta^{+}\nu)^{18}{\rm F}(p,\alpha)^{15}{\rm O}.$$
But if $^{19}{\rm Ne}$ captures a proton by the 
reaction $^{19}{\rm Ne}(p,\gamma)^{20}{\rm Na}$, the nuclear reactions
go out of the hot CNO cycle loop, and follow the rp process.
This process burns hydrogen via successive proton
capture and $\beta$ decays, and produces a large range of heavy nuclei
\citep{Schatz1999, Schatz2001}. The long series of $\beta$ decays makes these
bursts typically much longer ($\sim 100$ sec) than the helium bursts.\\
{\it Regime 4}: finally at a very high accretion rate (comparable to the Eddington limit),
the helium burning temperature sensitivity becomes weaker than the cooling
rate's sensitivity \citep{Ayasli1982, Taam1996}. Therefore in this
regime, the stable burning sets in, and the thermonuclear bursts do not occur.
However, recently \citet{CooperNarayan2006} have shown that the bursts 
should not occur even if the accretion rate is more than 30\% of the Eddington
accretion rate, in case $^{15}{\rm O}(\alpha,\gamma)^{19}{\rm Ne}$ reaction rate 
is lower than usually assumed.
This appears to agree with observations \citep{CooperNarayan2006}.
Finally, \citet{CooperNarayan2007a} have suggested that for rapidly 
spinning neutron stars, burning instability sets in (i.e., bursts
ignite) preferentially near the equator at low values of 
accretion rate and off of the equator at higher accretion rates. 

Apart from usual type-I X-ray bursts, 
bursts of longer durations have been observed from several
sources (e.g., \cite{Kapteinetal2000, intZandetal2002, intZandetal2005}).
They have typical durations of $\approx 30$ minutes and
energies $\approx 10^{41}$ ergs. These bursts are known as 
long bursts. Even longer and more energetic thermonuclear bursts,
viz. superbursts, have been observed from a few
neutron star LMXBs (\citet{Cornelisse2000, Kuulkers2004,
Zand2004, Kuulkers2005, Keek2008} and references therein). 
The typical released energy, recurrence time and decay time of these
bursts are $\sim 10^{42}$ ergs, $\sim$ years and $\sim 1-3$ hours
respectively \citep{CummingBildsten2001, StrohmayerBrown2002,
Cooper2009}.
Thermally unstable $^{12}{\rm C}$ fusion \citep{WoosleyTaam1976,
TaamPicklum1978, BrownBildsten1998} at a column
depth of $\sim 10^{12}$ gm cm$^{-2}$ may cause the superbursts
\citep{CummingBildsten2001, StrohmayerBrown2002}.
But this explanation has a problem for H/He accretors, because 
only a small amount of carbon is expected
to remain after the burning of H and He via the rp-process. 
Although several solutions
of this problem have been proposed by \citet{CummingBildsten2001};
\citet{Cooper2006}; \citet{Cooper2009}, more detailed theoretical 
studies are required to understand the origin of superbursts.

The study of thermonuclear bursts provides a unique opportunity to 
understand some aspects of extreme physics. 
In particular, these bursts can be very useful to constrain the
neutron star parameters (see later sections), 
and hence to understand the nature of the stellar core matter at 
supranuclear densities (see \S~1).
However, in order to use these bursts as a reliable tool, they
have to be understood well (see, for example, \cite{Woosleyetal2004,
Weinbergetal2006a, Weinbergetal2006b,
Linetal2006, Hegeretal2007a, 
CooperNarayan2007b, Fiskeretal2008} 
for recent burst model calculations). One of the poorly explored aspects
of these bursts is the thermonuclear flame spreading on the neutron 
star surfaces. It is expected that the accreted matter quickly spreads
all over the neutron star surface \citep{Inogamov1999}. Therefore a burst may
ignite at a certain location on the stellar surface
\citep{CooperNarayan2007a, Maurer2008}, and then the 
burning region may expand to cover the entire surface 
\citep{Fryxell1982, Spitkovsky2002}.
The study of this flame spreading is not only essential
to understand the bursts, but also important to use these
bursts as a tool to probe the neutron star parameters 
(\S~\ref{BurstOscillationMethod}). However,
the theoretical modeling of thermonuclear flame spreading
including all the main physical effects is extremely difficult, and only
recently \citet{Spitkovsky2002} have provided some
insights in this field. Although these authors have ignored several physical
effects (magnetic field, strong gravity, etc.) in their model, they have
considered the effects of the Coriolis force, which is important as the
bursting neutron stars are rapidly spinning (typical spin frequency $\sim 300-600$ Hz).
The observational study of flame spreading is also no less difficult than
the theoretical study. This is because, this spreading mostly happens
within $\approx 1$ sec during the burst rise, and in order to study it
observationally, one needs to detect and measure the evolution of the burst
spectral and timing properties during this short time. 
Therefore, although the observational indications of flame spreading were 
found about a decade ago (e.g., \cite{Strohmayer1997}), 
only recently such spreading could be confirmed
and measured for a few bursts by analyzing the {\it RXTE} PCA data
\citep{BhattacharyyaStrohmayer2005, BhattacharyyaStrohmayer2006a,
BhattacharyyaStrohmayer2006b, BhattacharyyaStrohmayer2006c,
BhattacharyyaStrohmayer2007a, BhattacharyyaStrohmayer2007c}.
These findings provide the motivation for rigorous theoretical studies
including all the main physical effects, as well as for detailed
analysis of the burst rise data.

\subsubsection{Why are Thermonuclear X-ray Bursts Useful?}\label{WhyareThermonuclearX-rayBurstsUseful}

Thermonuclear X-ray Bursts can be very useful to constrain the neutron star
parameters due to the following reasons.\\
(1) They originate from the neutron star surfaces. Therefore, the 
stellar parameters (e.g., mass, radius, spin frequency) influence
the burst spectral and timing properties. Hence the modeling of
these properties can be used to measure several 
neutron star parameters simultaneously (see later).\\
(2) Various burst phenomena, such as photospheric radius expansion,
thermonuclear flame spreading, etc., can be useful to constrain
the neutron star parameters (see later).\\
(3) Relatively low magnetic fields ($\sim 10^7-10^9$ G) of the bursting
neutron stars simplify the modeling of the surface emission.\\
(4) The surface emission during a burst is typically about 10 times 
brighter than the rest of the X-ray emission. Hence the surface
emission can be isolated from the total emission relatively
easily, and can be used to constrain the neutron star parameters with
confidence. Note that this is not true for the non-burst emission, for
which it is difficult to distinguish between the surface and the non-surface
emissions with sufficient accuracy.\\
(5) The $\sim 10$ times higher intensity of bursts, relative to the non-burst
intensity, gives a significantly higher signal-to-noise ratio.\\
(6) Many bursts can be observed from the same neutron star. The values of 
neutron star mass, radius and spin frequency, which affect the burst properties, 
do not change from one burst to another for a given source. 
Therefore, the joint analysis of 
many bursts, keeping the model values of these stellar parameters 
same for all the bursts, can be very useful to constrain these parameters. This
technique is demonstrated in \citet{Bhattacharyyaetal2005} 
(see also \S~\ref{SummaryoftheMethodsUsingThermonuclearX-rayBursts}).\\
(7) Thousands of thermonuclear X-ray Bursts have been observed from about 
80 sources (\cite{Liu2007}; see also \cite{Galloway2008} for a catalog of
{\it RXTE} PCA bursts). Therefore, a large amount of burst data
is already available. 

\subsubsection{Continuum Spectrum Method}\label{ContinuumSpectrumMethod}

\begin{figure}[h]
\begin{center}
\includegraphics*[width=8cm,angle=0]{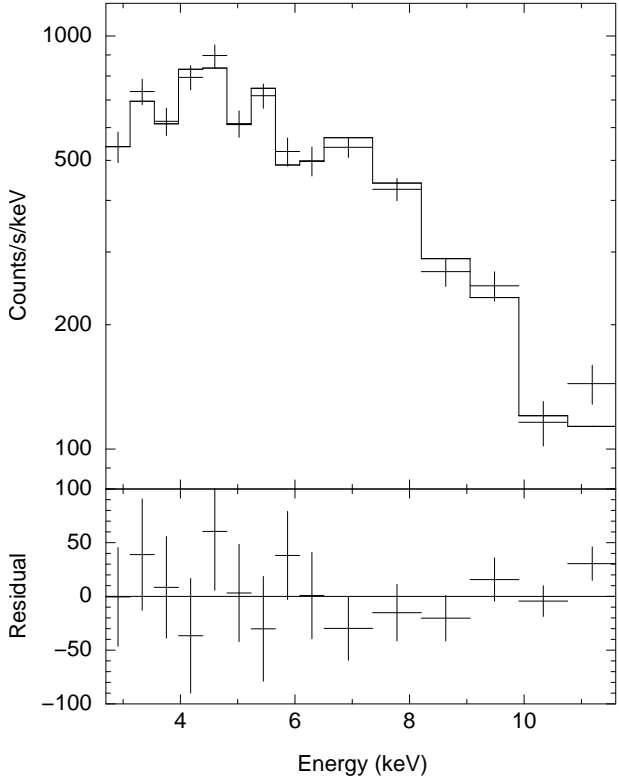}
\end{center}
\caption{This figure shows the fitting of a thermonuclear burst
spectrum with an absorbed blackbody model. The spectrum is for 
a 1 sec segment of a burst observed from a neutron star LMXB 4U 1636-536
with {\it RXTE} PCA. Since PCA cannot reliably measure the
absorption due to the Galactic neutral hydrogen because such
absorption mostly happens outside the PCA band, we fixed the
neutral hydrogen column density $n_{\rm H}$ to the estimated
value ($3.58\times10^{21}$ cm$^{-2}$) in the source direction for the fitting.
The upper panel shows the data points with error bars, and the best-fit
model, while the lower panel shows the (data$-$model). This figure
demonstrates that the continuum spectra of thermonuclear X-ray 
bursts are typically well described with a blackbody model
(\S~\ref{ContinuumSpectrumMethod}).
\label{spec_BB_fit}}
\end{figure}

The continuum spectra of thermonuclear X-ray bursts are typically
quite well described with a blackbody model (\citet{Swank1977,
Hoffmanetal1977b, Galloway2008}; see also Figure~\ref{spec_BB_fit}). 
The temperature of this blackbody
increases rapidly as the accumulated matter burns during the burst intensity rise,
and then decreases relatively slowly as the neutron star surface cools down 
during the burst decay. If the entire neutron star surface emits
like a blackbody of the same temperature at a given time, then the inferred
radius at infinity ($R_{\infty}$) of the neutron star can be obtained from the relation:
\begin{eqnarray}
R_{\infty} = (F_{\infty}/\sigma T^4_{\infty})^{1/2}d.
\label{BBrad1}
\end{eqnarray}
Here $F_{\infty}$ is the observed bolometric flux, $T_{\infty}$
is the fitted blackbody temperature, $d$ is the source
distance and $\sigma$ is the Stefan-Boltzmann constant. This provides
one of the very few ways to measure the neutron star radius.
Note that the radius measurement is typically more difficult than the
measurements of the neutron star mass and spin frequency. This
is because, whereas the gravitational effect of mass influences the
observed motion of the companion star in a binary system, and the observed
flux may be modulated by the neutron star spin, the radius exhibits
no such effects. As a result, although the mass and the
spin frequency of several neutron stars have been measured
\citep{ThorsettChakrabarty1999, LambBoutloukos2008},
accurate measurements of the radius are challenging.

The burst continuum spectrum method was used to constrain the
neutron star radius soon after the discovery of these bursts
(e.g., \cite{Paradijs1978, Goldman1979,
Paradijs1979, Paradijs1982, Marshall1982,
ParadijsLewin1986, FujimotoTaam1986, Sztajnoetal1987,
Gottwald1987}). 
However, these efforts were not too successful because
of the following systematic uncertainties.\\
(1) Equation~\ref{BBrad1} assumes that the entire stellar surface emits
during the burst. If that is not true, this equation
will underestimate the value of the neutron star radius.\\
(2) The lack of knowledge of the source distance introduces an
uncertainty in the neutron star radius value inferred from this method.\\
(3) The surface gravitational redshift ($1+z$) makes the
observed radius $R_{\infty}$ and the actual neutron star
radius $R_{\rm BB}$ different by an unknown factor. \\
(4) Although a blackbody model fits the observed burst spectrum 
well, this spectrum cannot be the blackbody originated
from the burning layer. This is because the 
scattering of photons by the electrons and the frequency 
dependence of the opacity in the neutron star
atmosphere hardens the energy spectrum \citep{Londonetal1984,
Londonetal1986, SyunyaevTitarchuk1986,
EbisuzakiNakamura1988, Titarchuk1994,
Madejetal2004, Majczynaetal2005}. 
Such a hardening shifts the burning layer blackbody spectrum
towards the higher energies, and changes its shape
slightly. This modified spectrum is called a ``diluted blackbody", and its
temperature ($T_{\rm col}$, i.e., the color temperature) is related to
the actual blackbody temperature of the burning layer
($T_{\rm BB}$) by an unknown color factor $f$. Consequently, the observed
radius is lower than the actual radius by the square of the
color factor. As a result,
the actual temperature ($T_{\rm BB}$) and radius ($R_{\rm BB}$)
are related to the observed values by the following relations
(including the effects of the gravitational redshift; \cite{Sztajno1985}):
\begin{eqnarray}
T_{\rm BB} = T_{\infty}.(1+z)/f;
\label{BBtemp1}
\end{eqnarray}
\begin{eqnarray}
R_{\rm BB} = R_{\infty}.f^2/(1+z);
\label{BBrad2}
\end{eqnarray}
where the surface gravitational redshift
$1+z =$ $(1-2GM/Rc^2)^{-1/2}$ for a non-spinning neutron star.
Here, $R/M$ is the neutron star radius-to-mass ratio,
$G$ is the gravitational constant and $c$ is the speed of light in vacuum.

\begin{figure}[h]
\begin{center}
\hspace*{-0.8cm}
\includegraphics*[width=10cm,angle=0]{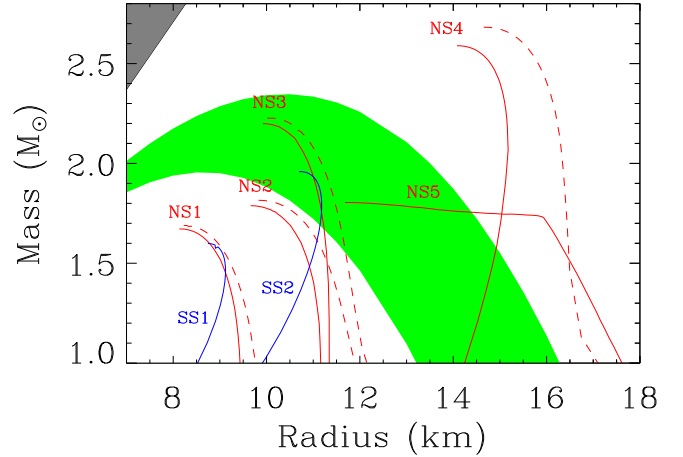}
\end{center}
\caption{
This figure shows the $M-R$ space of neutron stars with the curves
corresponding to a few representative EoS models 
(same as Figure~\ref{eosfirst}). The green patch
shows the allowed $M-R$ space assuming $R_{\rm BB} = R$ and
$1+z = (1-2GM/Rc^2)^{-1/2}$ in equation~\ref{BBrad2}, and
for $R_\infty.f^2$ in the range $15$ km $-$ $18$ km (see 
\S~\ref{ContinuumSpectrumMethod} for details). 
This range should be reasonable, because $R_\infty$ is often 
found in the $10$ km $-$ $20$ km range, and $f$ can be within 1.2 and 1.8 
according to \citet{Majczynaetal2005}.
This example figure demonstrates
how the thermonuclear burst continuum spectra can be 
used to constrain the neutron star mass and radius, and hence the EoS models.
\label{eosr}}
\end{figure}

As mentioned earlier, thermonuclear bursts provide one of the very 
few promising methods to measure the neutron star radius.
Therefore, although this method has a few systematic
uncertainties, it is worthwhile to try
to improve it by removing or at least reducing these uncertainties.
Here we briefly mention a few plausible ways to reduce the uncertainties.\\
(1) Even when a part of the stellar surface burns, 
$R^2_{\infty}$ should be proportional to,
and hence is a measure of the burning region area. Hence,
the uncertainty \#1 should be reduced if the burning region area can
be estimated independently; for example, by fitting the phase-folded
burst oscillation light curves with a physical model 
(see \S~\ref{BurstOscillationMethod}).\\
(2) In order to reduce the uncertainty \#2, one needs to measure the 
source distance by an independent method.
This is possible if the source is in a globular cluster 
(e.g., \cite{Ozeletal2009}); because the distances of these clusters
can typically be measured with sufficient accuracy. 
The source distance can also roughly be measured
from the flux of a photospheric radius expansion burst
(see \S~\ref{PhotosphericRadiusExpansionBurstMethod}).\\
(3) The surface gravitational redshift $1+z$ depends on the 
stellar parameters. As mentioned earlier, for a 
non-spinning neutron star, $1+z = (1-2GM/Rc^2)^{-1/2}$.
Therefore, the uncertainty \#3 can be reduced 
if the neutron star radius-to-mass ratio is constrained. 
Several observational features of 
neutron star LMXBs provide independent methods to measure this ratio
(for example, see \S~\ref{SpectralLineMethod};
\S~\ref{BurstOscillationMethod}; \S~\ref{BroadRelativisticIronLineMethod}).
However, if the uncertainties \#1, \#2 and \#4 can be sufficiently
reduced, it is possible to constrain the neutron star $M-R$ space
effectively, even without measuring the $1+z$. This is shown in 
Figure~\ref{eosr}, and can be understood from equation~\ref{BBrad2}
using $1+z = (1-2GM/Rc^2)^{-1/2}$.\\
(4) The color factor $f$ depends on the neutron star surface
gravity $g$, the actual blackbody temperature $T_{\rm BB}$ and the chemical
composition in the stellar atmosphere \citep{Madejetal2004,
Majczynaetal2005}. Therefore,
in order to measure $f$ and to reduce the uncertainty \#4, one needs to
estimate these three parameter values,
and model the atmosphere for these values.
While the gravity $g$ depends on the neutron star mass and radius
(see equation~\ref{gravity1}; see \S~\ref{MillihertzQuasi-periodicOscillationMethod}
and Figure~\ref{eos_g} for a plausible method to measure $g$),
the latter two parameter values can be estimated from
a realistic modeling of the thermonuclear bursts and
the chemical composition of the burning layer.
The determination of the composition may be attempted using the following ways. 
(a) The observed duration of the burst, as well as the detailed modeling
of its light curve and other properties may be useful to
determine the chemical composition (see \S~\ref{WhatisaThermonuclearX-rayBurst};
also see \cite{Woosleyetal2004, Hegeretal2007a}.
(b) The observed nature of the companion star can be used
to constrain the chemical composition of the accreted matter.
Note that both the preburst burning layer and the preburst atmosphere consist
of this matter, although the former may additionally contain ashes from the previous bursts. 
In case the companion is not directly observable, the measured binary orbital 
period may be useful to probe its nature. For example, if the 
system is an UCXB (orbital period $< 80$ min; \cite{zand2008}), the companion and hence
the accreted matter should be hydrogen deficient and heavier element 
rich (see \S~\ref{WhatisaThermonuclearX-rayBurst}).
(c) The observed atomic spectral lines from the non-burst emission can
also be useful to estimate the composition of the accreted matter.
Such lines are normally observed from high-inclination neutron
star LMXBs (e.g., \cite{Jimenez2003, trigoetal2006}).
Therefore, the detailed observational and theoretical studies
of burst and non-burst emissions can, in principle, constrain
the burst color factor $f$. However, note that $f$ is expected
to change, as $T_{\rm BB}$ and plausibly the atmospheric chemical 
composition evolve during a burst, and hence one needs to
estimate $f$ vaules throughout the burst.

\begin{figure}[h]
\begin{center}
\hspace*{-0.9cm}
\includegraphics*[width=10cm,angle=0]{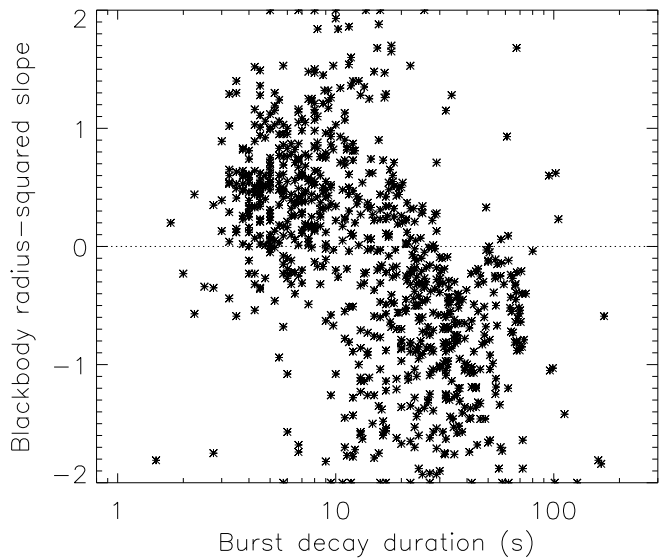}
\end{center}
\caption{Slope of burst blackbody radius-squared 
($R_{\infty}^2 \propto$ inferred burning area)
during burst decay vs. burst
decay duration. Each point corresponds to one burst, and 877
bursts from 43 sources have been shown.
This figure indicates a strong anticorrelation 
(significant at the $< 10^{-45}$ level) between the 
rate of change of the inferred burning area
and the decay duration, which can be useful to make the 
continuum burst spectrum method reliable (see \S~\ref{ContinuumSpectrumMethod}). 
For the exact definition of the radius-squared slope, see \citet{Bhattacharyyaetal2009}.
\label{burstcor}}
\end{figure}

The burning region is normally expected to cover the entire stellar 
surface during the burst decay (see \S~\ref{WhatisaThermonuclearX-rayBurst}).
If this is true, the uncertainty \#1 does not appear, and the continuum
spectrum of burst decay (rather than burst rise) should be more useful 
to measure the neutron star radius. But unfortunately, the inferred burning area
($\propto R^2_{\infty}$) can increase, decrease, or go up and down 
erratically in a given burst (see Figure~9 of \cite{Galloway2008}). 
This may be because of the actual change of $R^2_{\rm BB}$ and/or
the evolution of $f$ (see equation~\ref{BBrad2}) 
and/or any other reasons. Therefore, 
it is essential to understand the apparently erratic evolution of $R^2_{\infty}$
for a reliable measurement of neutron star radii using burst decay continuum spectra.
Very recently, \citet{Bhattacharyyaetal2009} have,
for the first time,
reported a pattern in the apparently erratic behaviour of the inferred burning area.
They found that the rate of change of the inferred area during decay
is anticorrelated with the burst decay duration, and this anticorrelation
is significant at the $<10^{-45}$ level (see Figure~\ref{burstcor}). 
These authors have suggested that 
the variations in $f$, and more specifically the variations in the
composition of the atmosphere between bursts with long and short durations,
may explain the anticorrelation. 
Such a variation can be reasonable, because the duration of a burst depends 
on the chemical composition of the burning material (see 
\S~\ref{WhatisaThermonuclearX-rayBurst}). However, in order to verify this model, 
extensive theoretical study is required. But, whatever is the correct model,
the newly discovered pattern in the $R^2_{\infty}$ evolution will play an extremely 
important role to make the continuum spectrum method reliable.

\subsubsection{Spectral Line Method}\label{SpectralLineMethod}

\begin{figure}[h]
\begin{center}
\hspace*{-0.8cm}
\includegraphics*[width=10cm,angle=0]{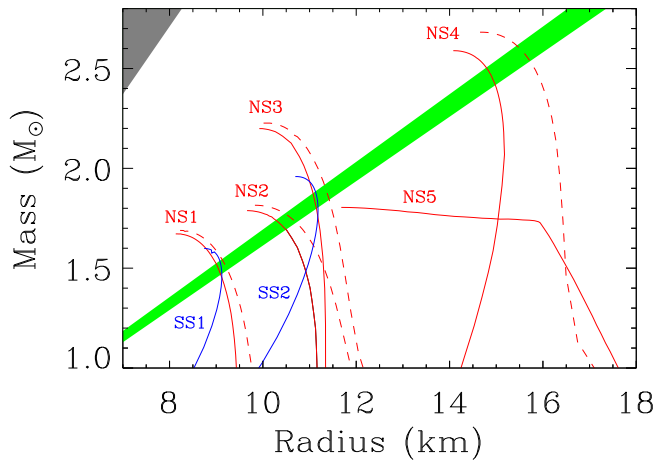}
\end{center}
\caption{
This figure shows the $M-R$ space of neutron stars with the curves
corresponding to a few representative EoS models 
(same as Figure~\ref{eosfirst}). The green patch
shows the allowed $M-R$ space for a reasonable range of
the radius-to-mass ratio ($Rc^2/GM = 4.0-4.2$).
This example figure demonstrates how the measured radius-to-mass ratio can be
used to constrain the neutron star $M-R$ space (see 
\S~\ref{SpectralLineMethod}; \S~\ref{BurstOscillationMethod};
and \S~\ref{BroadRelativisticIronLineMethod} for details).
\label{eosrbym}}
\end{figure}

Observation and identification of narrow atomic spectral lines from the surface of 
a neutron star (for example, during a thermonuclear burst) 
provide the cleanest way to measure the
stellar radius-to-mass ratio using the following formula (for the Schwartzschild
spacetime appropriate for a non-spinning neutron star):
$E_{\rm em}/E_{\rm obs} = 1+z = [1-(2GM/Rc^2)]^{-1/2}$ (\cite{OzelPsaltis2003,
Bhattacharyyaetal2006a}; see Figure~\ref{eosrbym}). 
Here, $E_{\rm em}$ is the emitted energy
of the line photons, and $E_{\rm obs}$ is the observed energy of these photons. 
However, if the surface lines are broad and asymmetric,
$E_{\rm obs}$ cannot be determined uniquely, and hence it is difficult to
measure $Rc^2/GM$ accurately using the above formula. Unfortunately,
neutron stars in LMXB systems typically spin with high frequencies
($\sim 300-600$ Hz), and hence the Doppler effect and the special relativistic
beaming are expected to make the surface lines broad and asymmetric.
Recently, \citet{Bhattacharyyaetal2006a} have suggested
a way to measure the stellar $Rc^2/GM$ using the above formula for such
broad lines. They have shown that using $\sqrt{E_{\rm l}E_{\rm h}}$
as $E_{\rm obs}$, one can measure $Rc^2/GM$ with less than 2\% error.
Here, $E_{\rm l}$ is the energy of the low energy end of the broad
line, and $E_{\rm h}$ is that of the high energy end. This shows that
surface lines can be used to accurately measure the stellar $Rc^2/GM$
even for LMXBs having rapidly spinning neutron stars.

\begin{figure}[h]
\begin{center}
\includegraphics*[width=8cm,angle=0]{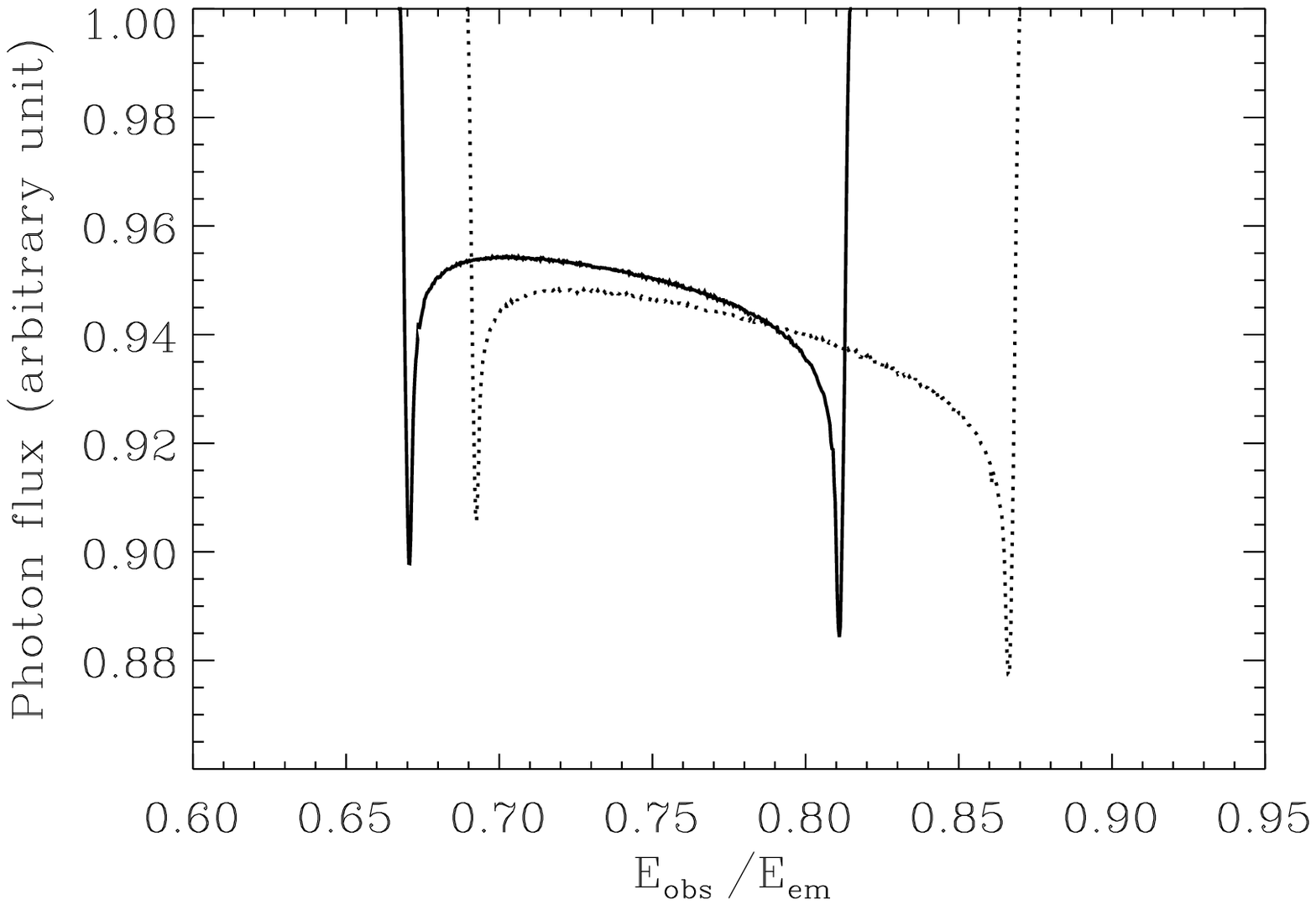}
\end{center}
\caption{Profiles of narrow absorption lines formed at the equatorial portion
of the neutron star surface as they would
appear to an observer at $i_{\rm spin} = 75^{\rm o}$, where $i_{\rm spin}$
is the observer's inclination angle measured from the stellar
spin axis (see \cite{Bhattacharyyaetal2006a}). Here, $E_{\rm em}$ is the emitted 
energy, $E_{\rm obs}$ is the observed energy, and the continuum is assumed to be
at a constant photon flux of 1 (in arbitrary unit). 
The solid profile is for the neutron star
parameter values $M=1.4M_\odot$ and $R=9$~km; and the dotted profile is for 
$M=1.5M_\odot$ and $R=11$~km. All other source parameter values are same,
including the stellar spin frequency ($= 400$~Hz). This figure shows that
the shape and the width of the surface spectral lines depend on the neutron star
parameters (\S~\ref{SpectralLineMethod}). 
\label{specline1}}
\end{figure}

A surface  absorption line originated from the entire stellar surface can have only one 
dip (or peak for an emission line; see Figure 1 of \cite{OzelPsaltis2003}).
On the other hand, if it originates from a portion of the surface,
especially near the equator, it can have two dips (or two peaks for
an emission line; \cite{OzelPsaltis2003, Bhattacharyyaetal2006a};
see also Figure~\ref{specline1}). In either case, its width and shape
depends on the neutron star parameter values (Figure~\ref{specline1};
see also \cite{OzelPsaltis2003, Bhattacharyyaetal2006a}).
Therefore, fitting the surface lines with physical models can be
useful to constrain additional (apart from $Rc^2/GM$) neutron star parameters.

Surface lines are likely to be detected and identified 
from thermonuclear X-ray bursts and hence from
neutron star LMXB systems. This is because: (1) the continuous accretion and
the radiative pressure may keep the line-forming heavy elements in the stellar atmosphere
for the time-duration required for the line detection; 
(2) nuclear burning ashes may be transported to the atmosphere by strong 
convection \citep{Weinbergetal2006a};
(3) the relatively low magnetic fields ($\sim 10^7-10^9$ Gauss)
of the neutron stars in LMXBs should make the line identification easier by keeping the
magnetic splitting negligible; 
and (4) the luminous bursts give good signal-to-noise ratio.
Moreover, the modeling of the surface lines detected from thermonuclear bursts
may be easier, because (1) magnetic field is relatively low; and (2) during the bursts,
 typically 90\% of the total emission originates from the
neutron star surface, and hence the surface spectral line should be largely free from the
uncertainty due to the other X-ray emission components, such as the accretion disk.

In fact, the only plausible observation of the neutron star surface atomic spectral
lines was from an LMXB EXO 0748-676 \citep{Cottametal2002}. These authors detected
two significant absorption features in the {\it XMM-Newton} 
Reflection Grating Spectrometers (RGS) energy spectra of the
thermonuclear X-ray bursts, which they identified as
Fe XXVI and Fe XXV $n=2-3$ lines with a gravitational redshift of $1+z = 1.35$.
In order to detect these lines, they had to combine the spectra of 28 bursts.
The absorption of some of the burst photons (originated in deep burning layers)
by the iron ions in the upper atmosphere might give rise to these redshifted
lines. \citet{Changetal2005} showed that the observed strength of the iron lines
could be produced by a neutron star photospheric metallicity, which was
$2-3$ times larger than the solar metallicity (see also \cite{Changetal2006}).
The identification of these two lines as the surface atomic spectral lines
is reasonable for the following reasons:
(1) the early burst phases (when the temperature was higher) showed the Fe XXVI
line, while the late burst phases (when the temperature was lower) exhibited the Fe XXV
line, which is consistent with the qualitative temperature dependence of the
iron ionization states; (2) the narrow lines were consistent with the relatively slow
stellar spin rate (45 Hz), reported by \citet{VillarrealStrohmayer2004}
(but see later); and
(3) both the lines showed the same surface gravitational redshift, and this redshift
value is reasonable  for the surface of a typical neutron star.
However, recent observations have argued against the neutron star surface
origin of these two lines.
(1) these lines were not significantly detected from the
later observations of EXO 0748-676 \citep{Cottametal2008}. 
However, such disappearence might be caused by the change in the photospheric conditions,
which might weaken the lines.
(2) \citet{Gallowayetal2009} have discovered burst oscillations at 
552 Hz from EXO 0748-676 with a significance of 6.2~$\sigma$. This 
argues that the neutron star spin frequency is 552 Hz 
(see \S~\ref{BurstOscillationMethod}). If this is true, the observed narrow 
spectral lines from EXO 0748-676 could originate from the stellar
surface only if the entire line forming region were very close to the
neutron star spin axis \citep{Bhattacharyyaetal2006a}.
Moreover, recently \citet{Balman2009} has suggested that the 45 Hz 
frequency reported by \citet{VillarrealStrohmayer2004} may be a
variability within the boundary layer.
Therefore, the surface origin of the observed absorption features from EXO 0748-676 
is not established yet. Besides, \citet{Kongetal2007} did not find surface
spectral lines from the combined {\it XMM-Newton} RGS spectra of 16
thermonuclear bursts from the neutron star LMXB GS 1826-24.

\subsubsection{Photospheric Radius Expansion Burst Method}\label{PhotosphericRadiusExpansionBurstMethod}

As mentioned in \S~\ref{WhatisaThermonuclearX-rayBurst}, for some of the strong bursts,
viz. photospheric radius expansion (PRE) bursts, 
the photospheric radiative pressure force exceeds the gravitational
force, and the photosphere is temporarily lifted off the neutron star surface.
During the expansion of the photosphere, the temperature decreases
and the burning area inferred from the continuum spectroscopy increases.
Subsequently, as the photosphere comes down, the temperature increases
and the inferred burning area decreases. After the photosphere settles on
the stellar surface at the touchdown time, it starts cooling down, and the 
temperature decreases again. As a result, PRE bursts can be identified by
two maxima in the temperature profile and a burning area maximum corresponding
to the temperature minimum (see Figure 9 of \cite{Galloway2008}). 
The amount of photospheric expansion can vary by a large extent
from burst to burst. Normally during the expansion, the X-ray flux does not 
change much. But for certain very powerful bursts the expansion can be so large
that the photospheric temperature may be shifted below the X-ray band,
and the burst may be divided into two parts: a precursor and a main part
\citep{Lewinetal1984, StrohmayerBrown2002}. For other somewhat
less powerful PRE bursts, a part of the flux may shift out of the X-ray
band making the burst intensity profile double-peaked 
\citep{StrohmayerBildsten2006}.

\begin{figure}[h]
\begin{center}
\hspace*{-0.8cm}
\includegraphics*[width=10cm,angle=0]{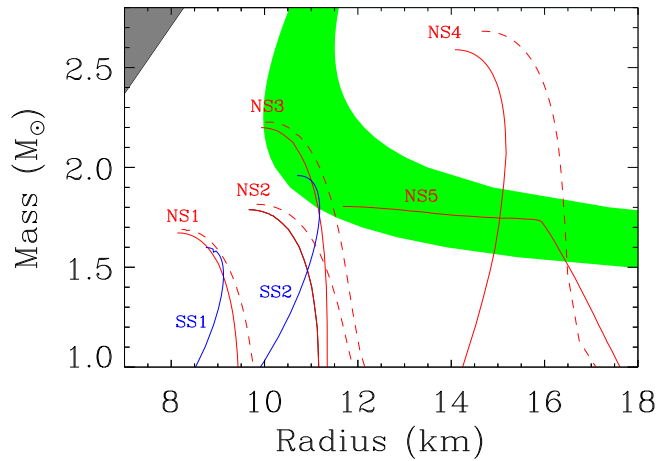}
\end{center}
\caption{
This figure shows the $M-R$ space of neutron stars with the curves
corresponding to a few representative EoS models 
(same as Figure~\ref{eosfirst}). The green patch
shows the allowed $M-R$ space using equation~\ref{Edd2} 
(see \S~\ref{PhotosphericRadiusExpansionBurstMethod} for details).
In this example figure we have ensured 
that the green patch passes through the the reasonable ranges of $M$ and $R$,
and this figure demonstrates how the photospheric radius expansion bursts can be
used to constrain the neutron star mass and radius, and hence the EoS models.
\label{eosedd}}
\end{figure}

The Eddington limit (see \S~\ref{WhatisaThermonuclearX-rayBurst}) on the
neutron star surface is given by
\begin{eqnarray}
\hspace{-0.8cm}
L_{\rm Edd} = (4\pi GMc/\kappa)(1-2GM/Rc^2)^{-1/2} = 4\pi R^2\sigma T^4_{\rm BB},
\label{Edd1}
\end{eqnarray}
where, $\kappa$ is the atmospheric opacity (e.g., \cite{StrohmayerBildsten2006}).
Here we have assumed that the entire stellar surface emits, and 
the surface gravitational redshift $1+z = [1-(2GM/Rc^2)]^{-1/2}$, 
which is appropriate for a non-spinning neutron star.
It may be reasonable to identify the observed touchdown flux as the
Eddington-limited flux \citep{Ozel2006}. Therefore, equations~\ref{Edd1} 
and \ref{BBtemp1} for the touchdown time can be used to constrain the neutron star 
$M-R$ space, if the entire stellar surface emits (see \cite{Gallowayetal2003}) 
and the parameters $\kappa$ and color factor $f$ are known. 
The flux observed at infinity corresponding to equation~\ref{Edd1}
is (e.g., \cite{Ozel2006})
\begin{eqnarray}
F_{\rm Edd} = (1/4\pi d^2)(4\pi GMc/\kappa)(1-2GM/Rc^2)^{1/2},
\label{Edd2}
\end{eqnarray}
where $d$ is the source distance.
Considering the observed touchdown flux as $F_{\rm Edd}$
\citep{Ozel2006}, equation~\ref{Edd2} can be used to constrain
the neutron star $M-R$ space (see Figure~\ref{eosedd}), 
if the entire stellar surface emits 
\citep{Gallowayetal2003} and the parameters $d$ and $\kappa$ are known. 
In order to estimate $\kappa$
and/or $f$, one needs to model the stellar atmosphere for a known 
chemical composition. The possible ways to determine this composition 
is discussed in \S~\ref{ContinuumSpectrumMethod}. Apart from the above methods,
the observed variation of the Eddington luminosity and the corresponding
change in the gravitational redshift caused by the photospheric expansion
may, in principle, be useful to determine the neutron star surface gravitational 
redshift (see for example, \cite{Damenetal1990, Paradijsetal1990}).
These and related methods have been used by many scientists in order
to constrain the neutron star mass and radius 
(e.g., \cite{FujimotoTaam1986, Ebisuzaki1987, Sztajnoetal1987, 
ParadijsLewin1987, ChevalierIlovaisky1990, Kaminkeretal1990, 
Damenetal1990, Paradijsetal1990, HaberlTitarchuk1995, 
Smale2001, Strohmayeretal1998, TitarchukShaposhnikov2002, 
Shaposhnikovetal2003, ShaposhnikovTitarchuk2004, Ozel2006}).
However, so far such measurements heve not been generally precise enough, because
of the systematic uncertainties primarily due to the unknown parameter values
mentioned above
(e.g., \cite{StrohmayerBildsten2006}).

This and other sections (e.g., \S~\ref{ContinuumSpectrumMethod}; 
\S~\ref{QuiescentEmissionMethod}) point out that the knowledge of
the source distance $d$ can be very useful to measure the
neutron star parameters. In 1978, \citet{Paradijs1978} suggested that the
photospheric expansion flux, which corresponds to the Eddington flux,
can be a ``standard candle", and hence a distance indicator (e.g., using
equation~\ref{Edd2}). In the next 30 years, scientists have tested
this idea by examining the variation of the photospheric expansion flux
from burst to burst, comparing the inferred distance to the known distance (whenever
available; e.g., for globular cluster sources), and in other ways
(e.g., \cite{Paradijs1981, Basinskaetal1984, Smale1998,
KuulkersvanderKlis2000, Kuulkersetal2003,
Gallowayetal2003, Gallowayetal2006,
GallowayCumming2006, Ozel2006, Gallowayetal2008};
Table 9 of \cite{Galloway2008}).
They have found that this method is affected by the uncertainties due to
unknown neutron star mass, surface gravitational redshift, photospheric
composition, etc., and hence the distance can typically be constrained
within a somewhat large range. 

\begin{figure}
\begin{center}
\hspace{-0.5cm}
\includegraphics*[width=9cm,angle=0]{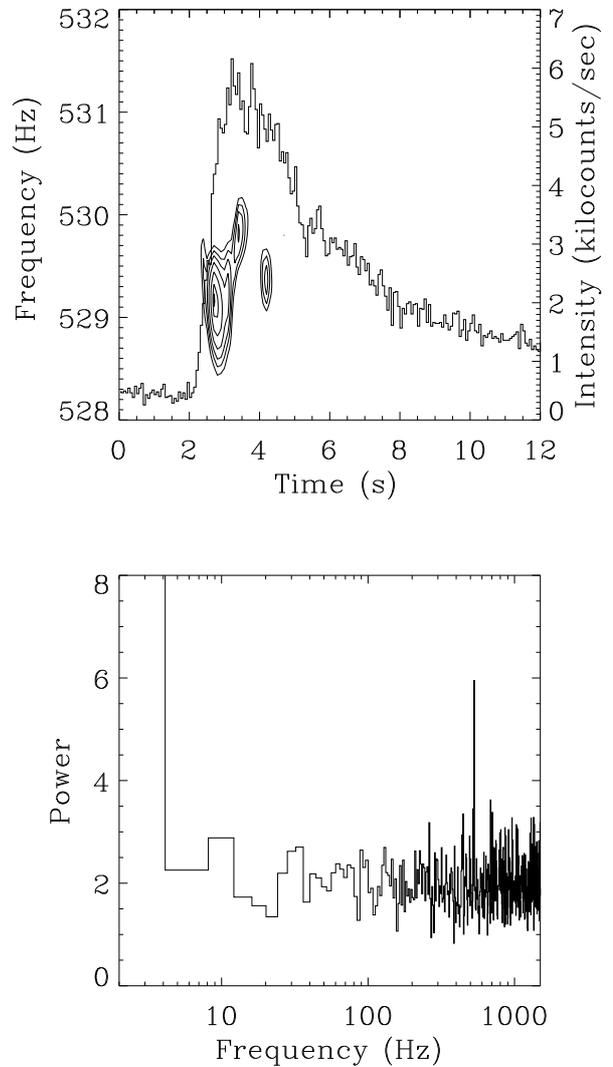}
\end{center}
\caption{Upper panel: X-ray intensity profile of a thermonuclear X-ray burst
from the neutron star LMXB 1A 1744-361 \citep{Bhattacharyyaetal2006b}.
Power contours around 530 Hz implies
that there are significant intensity variations (burst oscillations; 
\S~\ref{BurstOscillationMethod}) with $\approx 530$ Hz
frequency. Note that the shape of the power contours indicates a slight change
in frequency with time, which is common for burst oscillations.
Lower panel: The corresponding power spectrum. The
peak at $\approx 530$ Hz implies the burst oscillations with that frequency.
\label{burst_con_pow1_2}}
\end{figure}

\subsubsection{Burst Oscillation Method}\label{BurstOscillationMethod}

High frequency narrow timing features, viz. burst oscillations, 
have been detected from many thermonuclear X-ray bursts with the PCA 
instrument of {\it RXTE} \citep{StrohmayerBildsten2006}. They 
can be observed either in the burst rising part or in the decay part, or in
both. However, neither these oscillations appear in every
burst from a given source, nor they have been observed from every
neutron star LMXB \citep{Galloway2008}. 
Burst oscillations were first discovered from the LMXB 4U 1728-34 by
\citet{Strohmayeretal1996} in 1996. After that, this feature has been
detected from 20 other neutron star LMXBs (oscillations from three 
of them are not yet confirmed; see \cite{Bhattacharyya2007, 
LambBoutloukos2008, Gallowayetal2009, Wattsetal2009,
Markwardtetal2007a, Wattsetal2009a} and references therein).

Burst oscillations can be seen as a peak in the Fourier power spectrum computed
from a high time resolution light curve of either the entire burst or a part of
it (see Figure~\ref{burst_con_pow1_2}). The frequency of this feature can evolve during a 
burst (\cite{StrohmayerBildsten2006} and references therein). Normally
frequency shifts by less than 1\% of the mean frequency, but it can be somewhat
greater for some bursts. The frequency normally increases from a lower initial
value to an asymptotic value, but it has also been seen to decrease for some bursts
(see for example, \cite{Munoetal2002a}). 
It is believed that burst oscillations originate from an azimuthally 
asymmetric brightness pattern on the spinning neutron star surface.
Since in such a case the X-ray flux measured by the observer should
vary with the stellar spin frequency, the burst oscillation (asymptotic) frequency
is believed to be the stellar spin frequency. 
This model is established because of the following reasons.\\
(1) Burst oscillations originate from the
thermonuclear bursts which are neutron star surface phenomena. \\
(2) The oscillation (asymptotic) frequency is very stable
on timescales of years for a given source \citep{Munoetal2002a}.
This strongly suggests that burst oscillations are connected to the
stellar spin, which is also expected to be a stable property.\\ 
(3) The observed burst oscillation frequencies are mostly in the
range $\sim 300-600$ Hz. Such rapid spin rates are expected for neutron
stars in LMXBs, because of the accretion-induced angular momentum 
transfer (\S~\ref{Introduction}).\\
(4) Spin frequencies of five burst oscillation sources are independently
known from accretion-powered pulsations (see 
\S~\ref{Accretion-poweredMillisecondPeriodPulsations}).
For each of them the burst oscillation frequency matches the known
frequency well \citep{Chakrabartyetal2003, Strohmayeretal2003, 
Casellaetal2008, Wattsetal2009, Wattsetal2009a}.\\
The evolution of burst oscillation frequency may be caused by the
slow motion of the photosphere on the stellar surface. 
One reason for such a slow motion may be the angular momentum conservation
of contracting thermonuclear shells \citep{Strohmayeretal1997}.
In this scenario, a thermally expanded shell has a spin
frequency lower than the stellar spin frequency due to angular momentum 
conservation. As the shell contracts, its spin frequency increases 
and tends to the stellar spin frequency. However, this model cannot
explain a frequency shift roughly larger than 2 Hz 
\citep{StrohmayerBildsten2006} observed from several bursts (e.g.,
\cite{Gallowayetal2001, Wijnandsetal2001, Chakrabartyetal2003}).
Therefore, other physical reasons (e.g., involving magnetic field; 
\cite{BhattacharyyaStrohmayer2006c, 
HengSpitkovsky2009}) may contribute to the frequency evolution process.

Burst oscillations provide one of the two ways to measure the
spin rates of neutron stars in LMXB systems (see 
\S~\ref{Accretion-poweredMillisecondPeriodPulsations} for the other method).
In fact, this feature provided most of the known frequencies 
of such neutron stars. In addition, the amplitude and the shape of a phase-folded 
burst oscillation light curve can be very useful to constrain other stellar 
parameters. This is because the compactness ($GM/Rc^2$) and the spin-related 
surface speed ($\vartheta$) affect these light curves through the Doppler effect,
and special and general relativistic effects (including light bending).
The amplitude and asymmetry of phase-folded light curves are greater for
lesser $GM/Rc^2$ values (e.g., \cite{Pechenicketal1983, Strohmayeretal1998,
MillerLamb1998, Brajeetal2000, Weinbergetal2001}),
and hence can be used to constrain the neutron star radius-to-mass ratio.
Besides, $\vartheta$ introduces a pulse phase dependent Doppler shift in
the X-ray spectrum \citep{StrohmayerBildsten2006}, 
which can be used to estimate $\vartheta$. Note that
$\vartheta$ can be useful to constrain the neutron star radius,
for the known stellar spin frequency (measured from burst oscillations).

Several studies have been done to measure the neutron star properties
based on the modeling of burst oscillations. These include
an investigation of the amplitude and harmonic content of the
phase-folded light curves due to a point-like hot spot \citep{MillerLamb1998};
and the effects of expanding single hot spot or two hot spots \citep{Nathetal2002},
or single and antipodal hot spots of varying size \citep{Weinbergetal2001}.
Moreover, \citet{Cadeauetal2007} and \citet{Morsinketal2007}
studied the effects of the oblate shape of a rapidly spinning neutron star
on the phase-folded light curves.

Apart from XTE J1814-338, no other neutron star LMXB has exhibited 
significantly asymmetric (nonsinusoidal) phase-folded 
burst oscillation light curves (but see \cite{BhattacharyyaStrohmayer2005}).
The symmetric (sinusoidal) nature of the phase-folded light curves can be
attributed to the limitation of the detector capability and/or to 
one of the following reasons \citep{Munoetal2002b}: 
(1) if there is one hot spot on the neutron star, 
it must either be close to the spinning pole or cover nearly half the 
stellar surface; or (2) if there are two hot spots, they must form very near 
the spinning equator. Information about the
neutron star parameters can be extracted only from the amplitude of 
such a symmetric light curve, and not from its shape. Therefore, 
\citet{Bhattacharyyaetal2005} fitted the nonsinusoidal burst decay oscillation light curves
of XTE J1814-338 with a relativistic model appropriate for a single hot spot.
This model had five input parameters for a chosen EoS model and a known
neutron star spin frequency. These parameters were, (1) neutron star radius-to-mass
ratio ($Rc^2/GM$), (2) polar angle of the center of the spot, (3) angular 
radius of the spot, (4) a parameter which gave a measure of the beaming in the
emitter's frame, and (5) the observer's inclination angle measured from the
stellar spin axis. \citet{Bhattacharyyaetal2005} could determine a lower limit
of $Rc^2/GM$ at the 90\% confidence level, in spite of the systematics due to
the unknown values of the other parameters. 
\citet{Wattsetal2005} and \citet{WattsStrohmayer2006}
studied the variability and the energy dependence of the burst oscillation 
properties of XTE J1814-338, which may be useful to reduce these systematic uncertainties.

Since XTE J1814-338 is an accretion-powered millisecond pulsar 
(\S~\ref{WhatisanAccretion-poweredAccretion-poweredPulsar}),
and its modulations in the accretion and thermonuclear burst emission
are phase-locked with each other (see \cite{Wattsetal2008}),
its burst decay oscillation could very well originate from a hot spot.
But it is difficult to understand what gives rise to a hot spot
for a non-pulsar during burst decay. Therefore, several 
groups have attempted to model burst decay oscillations of non-pulsars 
using an asymmetric brightness pattern more complex than a hot spot. Such a complex
pattern may be caused by Rossby waves (or r-modes) in the surface layers
\citep{LeeStrohmayer2005, Heyl2005, PiroBildsten2006}, 
shear instabilities \citep{Cumming2005}, etc. Neutron star parameters
(e.g., $Rc^2/GM$) can, in principle, be constrained, even if burst oscillations
originate from surface modes (see \cite{LeeStrohmayer2005, Heyl2005}).
However, in practice, it is very difficult, because one needs to identify
the burst oscillations with the correct surface mode number.

Unlike burst decay oscillations of non-pulsars,
the oscillations during burst rise are likely to be caused by a hot spot (e.g.,
\cite{BhattacharyyaStrohmayer2007c, Wattsetal2009}). 
This hot spot is an expanding burning region, which can be modelled
more reliably than the burst decay oscillations, in order to 
constrain the neutron star parameters. A better
understanding of thermonuclear flame spreading 
(\S~\ref{WhatisaThermonuclearX-rayBurst}) will further increase
the reliability of neutron star parameter measurement using
burst rise oscillations. However note that, although an expanding 
hot spot is likely to cause the burst rise oscillations,
this hypothesis is yet to be comprehensively tested against the entire data set.

\subsubsection{Millihertz Quasi-periodic Oscillation Method}\label{MillihertzQuasi-periodicOscillationMethod}

In 2001, \citet{Revnivtsevetal2001} discovered milli-Hertz (mHz) period 
quasi-periodic oscillations (QPOs) from three neutron star LMXBs: 4U 1608--52,
4U 1636--536 and Aql X--1. This new class of timing feature has the following
properties: (1) the frequency varies between $0.007-0.009$ Hz (period 
$\approx 1.9-2.4$ minutes); (2) the
associated flux variations are at the few percent level; (3) the strength
is more at lower photon energies ($< 5$ keV; this is in contrast to other 
types of QPOs); (4) it appears in a narrow luminosity range around
$10^{37}$ ergs s$^{-1}$; and (5) it disappears after thermonuclear X-ray
bursts. Based on the theoretical research and observational findings of
\citet{Paczynski1983, Bildsten1993, Revnivtsevetal2001,  
YuVanderKlis2002, NarayanHeyl2003}, and others, 
\citet{Hegeretal2007b} proposed that the mHz QPOs correspond to the
marginal thermonuclear burning stability on the neutron star surfaces
close to the bounday between
{\it regime} 3 and {\it regime} 4 mentioned in \S~\ref{WhatisaThermonuclearX-rayBurst}.
Near such a bounday, oscillations are expected because the eigenvalues
of the system are complex \citep{Paczynski1983, NarayanHeyl2003}.
The narrow range of accretion rate corresponding to the bounday can explain
why mHz QPOs appear in a narrow luminosity range. \citet{Hegeretal2007b}
found that the period of the oscillations due to the marginal burning stability
should be $\approx \sqrt{t_{\rm thermal}t_{\rm accretion}}$ ($\approx 2$ min), where
$t_{\rm thermal}$ is the thermal timescale of the burning layer and 
$t_{\rm accretion}$ is the time to accrete the fuel at the Eddington
accretion rate. This theoretical period matches well with the observed
period, which is the main strength of the \citet{Hegeretal2007b} model.
However, note that a few aspects of mHz QPOs 
are still not understood \citep{Hegeretal2007b}.

\begin{figure}[h]
\begin{center}
\hspace*{-0.8cm}
\includegraphics*[width=10cm,angle=0]{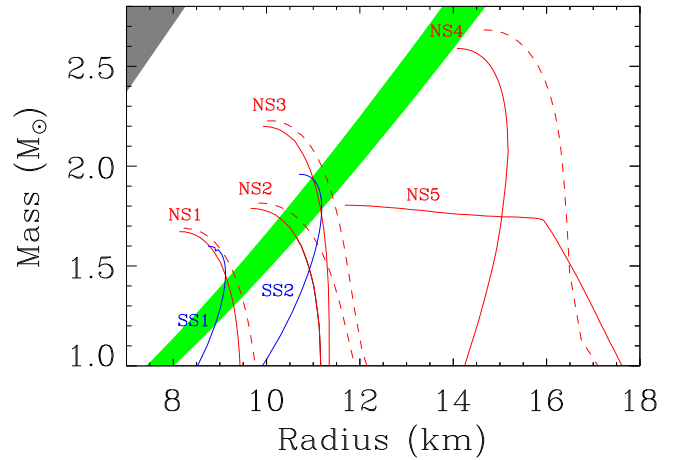}
\end{center}
\caption{This figure shows the $M-R$ space of neutron stars with the curves
corresponding to a few representative EoS models
(same as Figure~\ref{eosfirst}). The green patch
shows the allowed $M-R$ space
using equation~\ref{gravity1} for a reasonable range of
the surface gravity ($g = [2.6-3.1]\times10^{14}$ cm s$^{-2}$).
This example figure demonstrates how the measured surface gravity can be
used to constrain the neutron star $M-R$ space (see
\S~\ref{MillihertzQuasi-periodicOscillationMethod}).
\label{eos_g}}
\end{figure}

The theoretically proposed oscillation frequency is a strong function
of the surface gravity and the chemical composition of the accreted matter
(see Figure 9 of \cite{Hegeretal2007b}). Therefore, if the \citet{Hegeretal2007b} 
model is correct, and if the chemical composition can be determined (see
\S~\ref{ContinuumSpectrumMethod} for a discussion of plausible methods),
then mHz QPOs can be used to constrain the $M-R$ space of neutron stars
(see Figure~\ref{eos_g}).
This is because the surface gravity $g$ is a function of stellar mass and 
radius 
\begin{eqnarray}
g \simeq GM/R^2[1-2GM/Rc^2]^{-1/2}.
\label{gravity1}
\end{eqnarray}
Here we note that the 
marginal burning stability depends on the stellar spin rate \citep{Keeketal2009},
and hence the neutron star spin can also affect the mHz QPOs.
However, fortunately the stellar spin frequencies of all the currently known mHz QPO 
sources have meen measured \citep{LambBoutloukos2008}.

\citet{YuVanderKlis2002} found that kilohertz QPO (see 
\S~\ref{KilohertzQuasi-periodicOscillations}) frequency is anticorrelated
with the luminosity when mHz QPOs are observed. This is opposite
to the long-term trend. Therefore, mHz QPOs affect the kilohertz QPOs,
and hence can be useful to constrain the models of the latter. Such 
a constraint can be very helpful to measure the neutron star parameters
(see \S~\ref{KilohertzQuasi-periodicOscillationMethod}; 
\S~\ref{BroadRelativisticIronLineMethod}). \citet{Altamiranoetal2008}
found that mHz QPOs can also be used to predict the occurrence of 
thermonuclear bursts. Such a prediction may be helpful to observe 
more bursts, as the duty cycle of bursts is small.

\subsubsection{Joint Application of the Methods based on Thermonuclear X-ray Bursts}\label{SummaryoftheMethodsUsingThermonuclearX-rayBursts}


Thermonuclear X-ray bursts provide various methods to measure
the neutron star parameters (\S~\ref{ContinuumSpectrumMethod}; \S~\ref{SpectralLineMethod};
\S~\ref{PhotosphericRadiusExpansionBurstMethod}; \S~\ref{BurstOscillationMethod};
\S~\ref{MillihertzQuasi-periodicOscillationMethod}).
Figures \ref{eosr}, \ref{eosrbym} and \ref{eosedd} show how
some of these methods can be used to constrain the neutron star
$M-R$ space. Several of the burst properties, which provide
these methods, can be observed from a given
neutron star LMXB. For example, apart from the surface spectral lines,
all these burst features are observed from 4U 1636-536.
Similarly, all the burst features, except mHz QPOs, have
possibly been observed from EXO 0748-676. Therefore, joint
application of the burst methods can be very useful to constrain
the parameters of the same neutron star. For example, the continuum spectrum method
(\S~\ref{ContinuumSpectrumMethod}), the spectral line method (\S~\ref{SpectralLineMethod})
and the PRE burst method (\S~\ref{PhotosphericRadiusExpansionBurstMethod})
together may constrain the $M-R$ space effectively (e.g., the intersection
of the green patches of Figures \ref{eosr}, \ref{eosrbym} and \ref{eosedd};
see also Figure 1 of \cite{Ozel2006}). The first and the third of the
above methods, and the known source distance,
can also reduce the systematic uncertainties (e.g., \cite{Ozeletal2009}).
Moreover, \citet{BhattacharyyaStrohmayer2007a} have demonstrated that the
joint application of the continuum spectrum method and 
the burst oscillation method (\S~\ref{BurstOscillationMethod})
can constrain the neutron star parameters by eliminating the source
distance parameter.

A given neutron star LMXB can exhibit many bursts with various
properties (e.g., oscillations, photospheric radius expansion, etc.).
These bursts may be observed with different satellites (e.g.,
{\it RXTE}; {\it XMM-Newton}) at different times. 
The burst properties depend on neutron star and other system parameters.
Neutron star global parameter values remain unaltered for all these
observations, although some other source parameter values 
(e.g., the hot spot size for oscillations)
are expected to change. Here we briefly describe how these
observations can be combined within the framework of Bayesian statistics,
in order to constrain the neutron star parameters effectively
(see \cite{Bhattacharyyaetal2005}).
Let the posterior probability
density (which we determine by fitting the data with a model)
over the full set of model parameters $x_1...x_n$ be $p(x_1...x_n)$.
If we are interested in the confidence regions for a single parameter $x_k$, then
we integrate this probability distribution over the ``nuisance''
parameters $x_1...x_{k-1}$ and $x_{k+1}...x_n$: 
\begin{eqnarray}
q(x_k) = \int{\rm d}x_1...{\rm d}x_{k-1}{\rm d}x_{k+1}...{\rm d}x_np(x_1...x_n).
\label{Bayesian1}
\end{eqnarray} 
The maximum likelihood of $x_k$, as well as its confidence regions, can be
obtained from the probability function $q(x_k)$.
Using this method, one can obtain a tight joint constraint on a neutron star
global parameter (e.g., mass, radius, etc.) which does not vary from 
one observation to another, by integrating over the
nuisance parameters of all the observations.

\subsection{Accretion-powered Millisecond Period Pulsations}\label{Accretion-poweredMillisecondPeriodPulsations}

\subsubsection{What is an Accretion-powered Millisecond Pulsar?}\label{WhatisanAccretion-poweredAccretion-poweredPulsar}

Several neutron star LMXBs exhibit coherent fast (period $\sim 1$~millisecond)
brightness pulsations in their persistent X-ray emissions 
(e.g., non-burst, non-quiescent, etc.; \S~\ref{WhatisaThermonuclearX-rayBurst}; 
\S~\ref{WhatisaQuiescentEmission}; \cite{wijnands2005,  
Poutanen2008, Altamiranoetal2009a} and references therein). 
These pulsations can be detected as a narrow
peak in the power spectrum. In these sources, the accretion flow is channeled
by the neutron star magnetic field to the magnetic poles. The X-ray hot spots
on the neutron star created by this channeled flow rotate around the stellar
spin axis (if the magnetic axis and the spin axis are misaligned) and
give rise to the observed X-ray pulses. Note that one of the hot spots
is usually obsecured by the accretion disk.
Since the gravitational energy release from the accreted matter powers
the X-ray pulsations, these sources are called ``accretion-powered
millisecond pulsars (AMPs)".

In 1998, the first AMP SAX J1808.4-3658 was discovered with the {\it RXTE}
satellite \citep{WijnandsvanderKlis1998, ChakrabartyMorgan1998}.
This AMP exhibits significant pulsations at $\approx 401$ Hz. The data
revealed that (1) the low-energy pulses lag the high-energy pulses,
and (2) the pulse profiles are significantly asymmetric (nonsinusoidal) 
allowing a detection of even the second harmonic \citep{Cuietal1998}.
\citet{PoutanenGierlinski2003} suggested that both these properties
could be explained in terms of two spectral components: 
a `pencil'-like beamed blackbody emission and a `fan'-like Comptonized 
emission (from a radiative shock; but see also \cite{Hartmanetal2009}). 

Four years after the discovery of the first AMP, the second AMP
XTE J1751-305 was reported by \citet{Markwardtetal2002}. Since then
ten more AMPs have been discovered (\cite{wijnands2005, Altamiranoetal2009a,
Altamiranoetal2009b, Markwardtetal2009}; and references therein).
All these AMPs, including the first two, are transient sources 
(pulsations are detected only during the outbursts; \S~\ref{WhatisaQuiescentEmission}; 
\cite{Lambetal2008a}), and three of them show pulsations very occasionally
(see \cite{Wattsetal2009} and references therein). 
Here we do not describe the properties of these AMPs in detail, and
suggest the following references for the interested readers:
\citet{Gilfanovetal1998}; \citet{PsaltisChakrabarty1999};
\citet{BildstenChakrabarty2001}; \citet{Milleretal2003}; \citet{NelsonRappaport2003};
\citet{wijnands2005}; \citet{vanStraatenetal2005}; 
\citet{PoutanenBeloborodov2006}; \citet{vanderKlis2006};
\citet{Wattsetal2008}; \citet{Poutanen2008};
\citet{Cackettetal2009a}; etc.
Note that, although many properties of AMPs are broadly
understood, it is still a matter of debate why only a few neutron star
LMXBs are AMPs (e.g., \cite{Titarchuketal2002, Lambetal2008a, Ozel2009}).
Study of the intermittent pulsation sources, among other things, may be useful to solve
this problem (e.g., \cite{Lambetal2008b, Wattsetal2009}).

\subsubsection{Accretion-powered Millisecond Pulsation Method}\label{Accretion-poweredMillisecondPulsationMethod}

Accretion-powered pulsation frequency is the neutron star spin frequency,
because such pulsations are caused by the periodic motion of a hot spot 
due to the stellar spin (see ~\ref{WhatisanAccretion-poweredAccretion-poweredPulsar}).
These pulsations are more coherent than burst oscillations 
(\S~\ref{BurstOscillationMethod}), and, 
unlike the burst oscillation frequency, the pulsation frequency 
does not evolve appreciably in a short time. Therefore, the pulsation method
gives a more accurate measurement of the spin frequency.

The phase-folded accretion-powered pulsation light curves are affected by
the Doppler effect, and the special and general relativistic effects.
Therefore, similar to the burst oscillations, the amplitude and the shape of a pulsation
light curve can also be very useful to measure the neutron star and other system
papameters (see \S~\ref{BurstOscillationMethod} for the burst oscillation method).
Various groups attempted to constrain the stellar parameters by fitting
the pulsation light curves with appropriate relativistic models
(e.g., \cite{PoutanenGierlinski2003, PoutanenGierlinski2004, 
Leahyetal2008, Leahyetal2009}). However, similar to
the burst oscillation method (see \S~\ref{BurstOscillationMethod}),
the systematic uncertainties due to the unknown system parameters
(such as, observer's inclination angle, size and polar angle of the 
hot spot, etc.) hinder an accurate measurement. Besides, additional
systematics due to the uncertainties in the spectral properties of
various X-ray emitting components make the pulsation method less
reliable (see the next paragraph).

Both burst oscillations and accretion-powered pulsations originate
from azimuthally asymmetric brightness patterns on the spinning
neutron star surfaces. Therefore, in principle, similar methods can be used
to extract information from them. However, in practice, these two
methods are somewhat different from each other because of the 
following reasons. \\
(1) Phase-folded burst oscillation light curves appear symmetric
or sinusoidal with the current observational capabilities
(see, \cite{Strohmayeretal2003} and \cite{BhattacharyyaStrohmayer2005} for exceptions).
However, the accretion-powered pulsation light curves can be nonsinusoidal, especially,
but not exclusively, at higher energies (e.g., \cite{Cuietal1998, Poutanen2008}).
Therefore, in principle, more information can be extracted from the
accretion-powered pulsation light curves (using both amplitudes and shapes; 
see \S~\ref{BurstOscillationMethod}). \\
(2) During a thermonuclear burst, most of the emission originates from the
neutron star surface, and the corresponding energy spectrum can be fitted
well with a blackbody model (see \S~\ref{ContinuumSpectrumMethod}).
Moreover, only burst emission contributes to the oscillations, and 
the gravitational energy release due to the accretion can be
separated from this burst emission (see \cite{Galloway2008}). These
make the modeling of phase-folded burst oscillation light curves simple
and reliable. But during accretion-powered pulsations, gravitational 
potential energy powers both the disk and the surface emissions. Moreover, the
surface emission, which gives rise to the pulsations, is at least 
partially processed (Comptonized) in the bulk motion of matter
(e.g., \cite{MastichiadisKylafis1992}) channelled by the stellar magnetic 
field, and/or in a radiative shock
at the neutron star surface \citep{PoutanenGierlinski2003}.
It is very difficult to observationally separate these various components of emission, 
as well as their contributions to the phase-folded pulsation light curves, from each other.
This makes the understanding of pulsation light curves difficult, and
reduces the reliability of the pulsation method. Rigorous theoretical
computation of both time-averaged and phase-resolved energy spectra
is required to solve this problem. For example, among other things,
one may need to perform the Monte Carlo simulation of the scattering
of photons in the bulk electron flow near the magnetic poles,
including the gravitational redshift and light bending effects.\\
(3) Unlike burst oscillations, accretion-powered pulsations can be useful 
to test the effects of accretion rate changes on the inner-edge radius
of the accretion disk, the size of the hot spot, etc. \citep{IbragimovPoutanen2009}. 
This can, in turn, be useful to estimate a few system parameters (e.g.,
neutron star magnetic field; \cite{IbragimovPoutanen2009}), and hence to make some of the
methods described throughout this review more reliable.

Finally, we note that other methods have been proposed in the literature to constrain
the neutron star $M-R$ space for AMPs \citep{Bhattacharyya2001}. 
\citet{Gilfanovetal1998} reported that the X-ray energy spectrum of 
SAX J1808.4-3658 had remained stable, when the X-ray luminosity
varied by two orders of magnitude. 
\citet{Bhattacharyya2001} (see also \cite{BurderiKing1998, Lietal1999}) 
found the upper limits to the mass and the radius of the neutron 
star using this observational result, and with the following
assumptions: (1) the neutron star magnetic field is dipolar;
(2) there is no ``intrinsic" pulse mechanism; and 
(3) the accretion flow is not centrifugally inhibited.

\subsection{Kilohertz Quasi-periodic Oscillations}\label{KilohertzQuasi-periodicOscillations}

\subsubsection{What are Kilohertz Quasi-periodic Oscillations?}\label{WhatareKilohertzQuasi-periodicOscillations}

\begin{figure}[h]
\begin{center}
\includegraphics[width=3.00in]{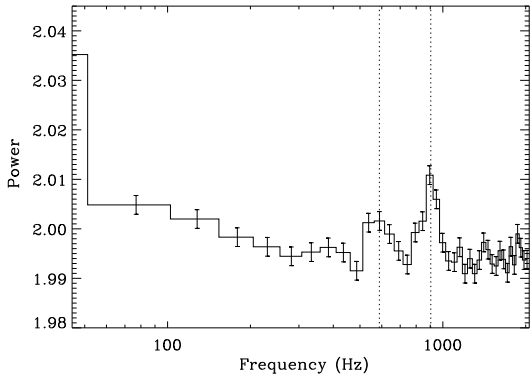}
\end{center}
\caption{Power spectrum of a neutron star LMXB showing a pair of kHz QPOs 
(see \S~\ref{WhatareKilohertzQuasi-periodicOscillations}).
The vertical dotted lines show the centroid frequencies of these QPOs.
\label{khzqpo2}}
\end{figure}

Neutron star LMXBs exhibit a zoo of timing features, both broad and narrow, 
in the Fourier power spectra (see \cite{vanderKlis2006} for an excellent review).
Many of these features are correlated with each other and with the spectral
states of the source (\cite{vanderKlis2006} and references therein). 
Among these features, high 
frequency quasi-periodic oscillations, or kilohertz (kHz) QPOs, are the
most important ones to constrain the neutron star parameters 
(see \S~\ref{KilohertzQuasi-periodicOscillationMethod}). These QPOs often occur
in a pair (Figure~\ref{khzqpo2}), and the twin peaks usually move together
in the frequency range $\sim 200-1200$ Hz in correlation with the source state.
The higher frequency QPO is known as the upper kHz QPO (frequency $= \nu_u$),
and the lower frequency QPO is called the lower kHz QPO (frequency $= \nu_l$).
The kHz QPOs were first reported in \citet{vanderKlisetal1996} and 
\citet{Strohmayeretal1996} (see also \cite{vanderKlis1998} for a historical 
account). As of now these QPOs have been observed from about 30 neutron star
LMXBs (see \cite{Altamiranoetal2009a, vanderKlis2006}). 

Although frequency ($\nu$), amplitude, quality factor ($\nu/\delta\nu$) and
other properties of kHz QPOs vary from source to source, and even for a 
given source depending on its intensity and spectral state, several
general trends of this timing feature could still be established.
Here we mention some of them which may be useful to understand these QPOs.
(1) For a given source, $\nu$ correlates well with the X-ray luminosity
$L_x$ on time scales of hours. But on longer time scales, and across sources,
the $\nu - L_x$ correlation does not exist, and similar $\nu$ values are
observed over two orders of magnitude in $L_x$ (e.g., \cite{vanderKlis1997}).
(2) It was thought that the frequency separation $\nu_u-\nu_l$ ($= \Delta\nu$) clusters 
around the neutron star spin frequency and the half of it (\cite{Wijnandsetal2003, 
vanderKlis2006} and references 
therein); although the recent works of \citet{MendezBelloni2007} and others have
suggested that $\Delta\nu$ is consistent with no dependence on spin.
(3) KHz QPO amplitudes increase with photon energy, and in similar energy bands,
the QPOs are weaker in the more luminous sources (\cite{Jonkeretal2001, 
vanderKlis2006} and references therein).
(4) The quality factor of lower kHz QPOs increases with the QPO frequency at
lower frequency values, and decreases at higher frequency values. But the 
quality factor of upper kHz QPOs increases with the frequency all the
way to the highest detectable frequencies \citep{Barretetal2006}.
(5) KHz QPO frequencies are correlated with the frequencies (or characteristic
frequencies) of many narrow and broad timing features from neutron
star LMXBs (see Figure 2.9 of \cite{vanderKlis2006}).

\begin{figure}[h]
\begin{center}
\hspace*{-0.5cm}
\includegraphics*[width=9.5cm,angle=0]{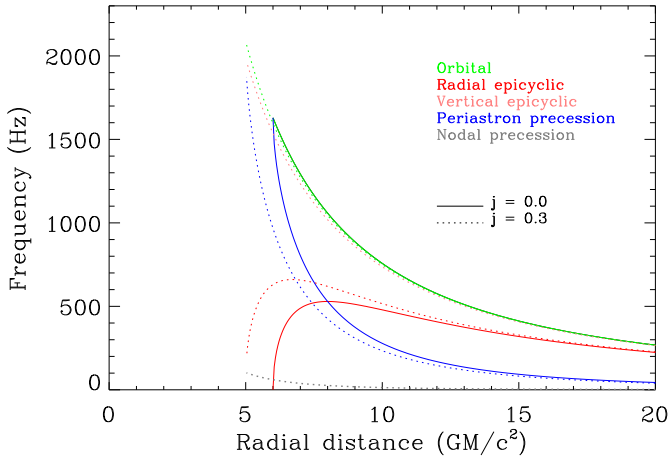}
\end{center}
\caption{This figure shows the radial profiles of various frequencies (color coded) 
of equatorial circular orbits in Kerr spacetime (see \S~\ref{WhatareKilohertzQuasi-periodicOscillations}).
Two angular momentum parameters ($j = 0.0$ (solid) and $j = 0.3$ (dotted)),
and neutron star mass $= 1.35 M_\odot$ are used. These frequencies may be useful to
understand the kHz QPOs (see \S~\ref{WhatareKilohertzQuasi-periodicOscillations}).
\label{frequency1}}
\end{figure}

What causes the kHz QPOs is not known yet, although many models are available
in the literature. A discussion of all these models is out of the scope of
this review. Here, we will mention only a few models which will be used
in \S~\ref{KilohertzQuasi-periodicOscillationMethod} to demonstrate 
the plausible ways to measure neutron star parameters using kHz QPOs.
The most prominent property of a kHz QPO is its high frequency. Most
of the models have primarily attempted to explain this property.
This frequency is usually believed to be caused by the frequencies 
related to the fast motions close to a neutron star,
where the effects of general relativity is important. Let us first identify 
some of these frequencies, and write down their expressions for circular orbits
in the equatorial plane for Kerr spacetime. \\ 
(1) Orbital frequency around the neutron star 
\begin{eqnarray}
\nu_\phi = \nu_{\rm K} (1+j(r_g/r)^{3/2})^{-1},
\label{freq1}
\end{eqnarray}
where $\nu_{\rm K} = \sqrt{GM/r^3}/2\pi$. Here, $r$ is the radial distance from the 
center of the neutron star, $r_g \equiv GM/c^2$, and the angular momentum
parameter $j \equiv Jc/GM^2$, where $J$ is the stellar angular momentum. \\
(2) Radial epicyclic frequency (for infinitesimally eccentric orbits)
\begin{eqnarray}
\hspace{-0.55cm}
\nu_r = \nu_\phi (1-6(r_g/r)+8j(r_g/r)^{3/2}-3j^2(r_g/r)^2)^{1/2}.
\label{freq2}
\end{eqnarray}
(3) Vertical epicyclic frequency (for infinitesimally tilted orbits)
\begin{eqnarray}
\hspace{-0.25cm}
\nu_\theta = \nu_\phi (1-4j(r_g/r)^{3/2}+3j^2(r_g/r)^2)^{1/2}.
\label{freq3}
\end{eqnarray}
(4) Periastron precession frequency $\nu_{\rm peri} = \nu_\phi - \nu_r$. \\
(5) Nodal precession frequency $\nu_{\rm nodal} = \nu_\phi - \nu_\theta$. \\
Apart from these, the neutron star spin frequency $\nu_{\rm spin}$ may also
contribute to kHz QPOs. The radial profiles of the general
relativistic frequencies are shown in the Figure~\ref{frequency1}. This
figure helps to understand any model based on these frequencies. It also
clearly shows that the radial epicyclic frequency
cannot be $\nu_u$ or $\nu_l$, because some kHz QPOs are observed with 
frequencies much higher than the maximum possible value of $\nu_r$.
In many models, kHz QPO frequencies are explained in terms of 
the above mentioned general relativistic frequencies and $\nu_{\rm spin}$ 
(or beating between any two of them) at some preferred 
radii. A preferred radius may be determined by the disk flow structure
(depends on, for example, the accretion rate), the radius of the 
general relativistic innermost stable circular orbit (ISCO), or
the commensurabilities among the frequencies which may cause resonance.
For example, \citet{Milleretal1998} identified $\nu_u$ and $\nu_l$ 
with $\nu_\phi$ and a beating frequency ($\nu_{\rm beat}$) respectively
at the ``sonic" radius ($r_{\rm sonic}$). \citet{LambMiller2003}
modified this model maintaining $\nu_u = \nu_\phi (r_{\rm sonic})$, and
explaining $\nu_l$ by a beat-frequency interaction between orbital 
motion at $r_{\rm sonic}$ and an azimuthal structure at the spin-resonance
radius, defined by $\nu_{\rm spin} - \nu_\phi = \nu_\theta$.
\citet{StellaVietri1998,StellaVietri1999} identified $\nu_u$ with 
$\nu_\phi$ at the disk inner edge, and related $\nu_l$ and $\nu_h$
(frequency of a low-frequency QPO; see \cite{vanderKlis2006}) 
with $\nu_{\rm peri}$ and $\nu_{\rm nodal}$ respectively.
Various authors (e.g., \cite{KluzniakAbramowicz2001, AbramowiczKluzniak2001})
made use of the fact that $\nu_\phi$, $\nu_r$ and $\nu_\theta$
at particular radii have simple integer ratios or other commensurabilities with
each other or with $\nu_{\rm spin}$. They suggested that at these
radii resonances may occur which can show up as kHz QPOs.
Besides, many other authors have used the general relativistic 
frequencies in various ways to explain the properties of kHz QPOs
(e.g., \cite{Wijnandsetal2003, Leeetal2004,  
Zhang2004, Mukhopadhyay2009}).

Apart from the observed frequencies of kHz QPOs, one also needs 
to explain the modulation and decoherence mechanisms.
Some of the plausible methods have been discussed in \citet{vanderKlis2006}
and references therein. \citet{Mendez2006} proposed that, although the
kHz QPO frequencies are plausibly determined by the characteristic
disk frequencies (see the previous paragraph), the modulation mechanism
is likely associated to the high energy spectral component (e.g., 
accretion disk corona, boundary layer between the disk and the neutron star,
etc.). This is because the disk alone cannot explain
the large observed amplitudes, especially at hard X-rays where the contribution
of the disk is small. Detection and measurement kHz QPOs above $20-30$ keV
may be able to resolve this issue. In addition, observations of
sidebands, overtones, and very high frequency QPOs may be useful to
identify the correct kHz QPO model \citep{vanderKlis2006, Bhattacharyya2009}.
We will not discuss these aspects further. Rather
we note that the fluid dynamical simulation corresponding to any successful 
kHz QPO model must naturally give rise to the observed frequencies and other 
properties.

\subsubsection{Kilohertz Quasi-periodic Oscillation Method}\label{KilohertzQuasi-periodicOscillationMethod}

\begin{figure}[h]
\begin{center}
\hspace*{-0.8cm}
\includegraphics*[width=10cm,angle=0]{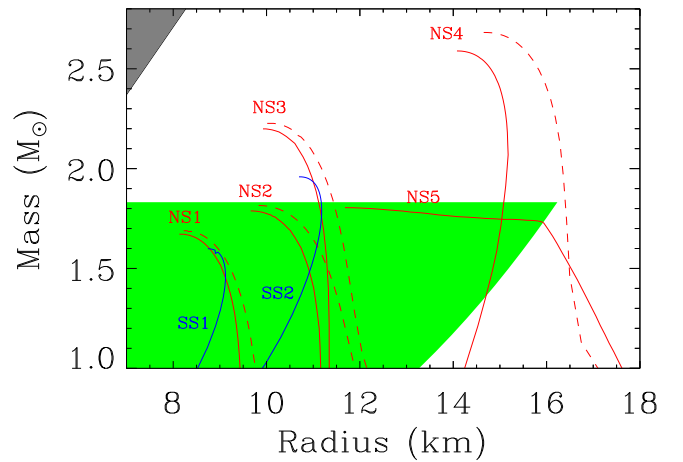}
\end{center}
\caption{This figure shows the $M-R$ space of neutron stars with the curves
corresponding to a few representative EoS models 
(same as Figure~\ref{eosfirst}). The green patch
shows the allowed $M-R$ space using equations~\ref{freq4} and \ref{freq5}, and
an upper kHz QPO frequency of 1200 Hz (\cite{Milleretal1998}; 
see \S~\ref{KilohertzQuasi-periodicOscillationMethod}).
This example figure demonstrates the potential of kHz QPOs to constrain the 
neutron star mass and radius, and hence the EoS models.
\label{eos_qpo}}
\end{figure}

Since no kHz QPO model can yet explain all the major aspects of 
this timing feature, currently it is not possible
to constrain the neutron star parameters using this QPOs with
certainty. However, all proposed models involve plasma motion in the
strong gravitational field around the neutron star, and with one
exception (photon bubbles; \cite{Kleinetal1996a, Kleinetal1996b})
suggest that the kHz QPOs originate in the disk \citep{vanderKlis2006}. 
Moreover, most models identify one of the kHz QPO frequencies (usually $\nu_u$; 
but can also be $\nu_l$) with the orbital motion at a 
preferred disk radius (\cite{vanderKlis2006}; see also 
\S~\ref{WhatareKilohertzQuasi-periodicOscillations}). 
If this is true, two reasonable
conditions can constrain the $M-R$ space \citep{Milleretal1998}:
\begin{eqnarray}
R \le r,
\label{freq4}
\end{eqnarray}
where $r$ is the radius of the orbit associated with 
$\nu_u$ or $\nu_l$ via equation~\ref{freq1}; and 
\begin{eqnarray}
r_{\rm ISCO} \le r,
\label{freq5}
\end{eqnarray}
where $r_{\rm ISCO}$ is the radius of the ISCO.
This is because the first condition gives a mass-dependent upper 
limit on $R$ via equation~\ref{freq1}; and the second condition
gives an upper limit on $M$: $M < c^3/(2\pi 6^{3/2}G\nu_\phi|_r)$
(for Schwarzschild spacetime). These constraints on $M-R$ space
are shown in Figure~\ref{eos_qpo}.

The general relativistic frequencies depend on the neutron star parameters
$M$ and $j$, and many models use these frequencies to explain kHz QPOs 
(\S~\ref{WhatareKilohertzQuasi-periodicOscillations}; see also
\cite{vanderKlis2006} and references therein). 
Therefore, in a more general sense the identification of any of the 
kHz QPO frequencies
with the beating, resonance, or any other combination of $\nu_\phi$,
$\nu_r$, $\nu_\theta$, $\nu_{\rm peri}$, $\nu_{\rm nodal}$ and
$\nu_{\rm spin}$ has the potential to constrain the neutron star
parameter space. Besides, the identification of $\nu_h$
with $\nu_{\rm nodal}$ (\S~\ref{WhatareKilohertzQuasi-periodicOscillations})
can be useful to constrain the stellar moment of inertia $I$,
because $\nu_{\rm nodal} = 8\pi^2\nu_\phi^2I\nu_{\rm spin}/Mc^2$ 
\citep{StellaVietri1998}. Finally, simultaneous observations of kHz
QPOs and broad relativistic iron lines have the potential to measure
neutron star parameters (see \S~\ref{BroadRelativisticIronLineMethod}).

\subsection{Broad Relativistic Iron Lines}\label{BroadRelativisticIronLines}

\subsubsection{What is a Broad Relativistic Iron Line?}\label{WhatisaBroadRelativisticIronLine}
A broad iron K$\alpha$ spectral emission line near 6 keV is observed from many
accreting supermassive and stellar-mass black hole systems (see \cite{Fabianetal2000, 
ReynoldsNowak2003, Miller2007}; and references therein). 
Although various models were proposed to explain this feature (e.g., 
\cite{MisraKembhavi1998, Titarchuketal2003, Reevesetal2004,
Titarchuketal2009}), 
the large width, the lower-energy wing and the asymmetry of such a line, as
well as the observed associated ``disk reflection'' spectrum strongly suggest that
this line originates from the inner part of the accretion disk, where it is
shaped by Doppler and relativistic effects (e.g., \cite{Tanakaetal1995,  
Fabianetal2000, Milleretal2001, Miniuttietal2007, 
Milleretal2009}). This relativistic line are
believed to be produced by the reflection of hard X-rays from the accretion disk.
The hard X-ray source may be anything from an accretion disk corona 
\citep{Fabianetal2000} to the 
base of a jet \citep{MarkoffNowak2004, Markoffetal2005}. 
A given incident X-ray photon may be Compton scattered by
free or bound electrons, or subject to photoelectric absorption followed by
either Auger de-excitation or fluorescent line emission \citep{Fabianetal2000}.
The strongest among the fluorescent spectral emission lines is the one for 
the $n = 2 \rightarrow n = 1$ transition of the iron atom (or ion).
This iron K$\alpha$ line originates at an energy between 6.4 keV and 
6.97 keV, depending on the iron ionization state \citep{ReynoldsNowak2003}.
The intrinsically narrow iron K$\alpha$ line is shaped by the following effects:
(1) the Doppler effect due to the motion of the line-emitting matter in the
inner part of the accretion disk broadens the line and makes it double-peaked;
(2) The general relativistic light-bending
effect also contributes to the formation of the two peaks;
(3) the transverse
Doppler shift and the gravitational redshift shift the line towards lower
energies; and 
(4) the special relativistic beaming reduces the lower-energy
peak and enhances the higher-energy peak making the line asymmetric 
(we encourage the readers to
see the Figure 3 of \cite{Fabianetal2000}).
Therefore, the relativistic iron line provides an ideal tool to probe
the spacetime and the flow and structure of accreted matter in the 
strong gravity region. The modelling of such a line can also be useful to 
constrain the radius of the ISCO (in the unit of black hole mass), 
which can be used to measure the
black hole spin parameter or angular momentum parameter ($j \equiv Jc/GM^2 \equiv a/M$;
$J$ and $M$ are the black hole angular momentum and mass respectively; 
see, for example, 
\cite{BrennemanReynolds2006, Milleretal2009}).
This is because $a/M$ has a one-to-one correspondence with the 
ratio of the ISCO radius to black hole mass for Kerr spacetime 
(see Figure 1a of \cite{Miller2007}).

\begin{figure}[h]
\begin{center}
\includegraphics[width=3.00in]{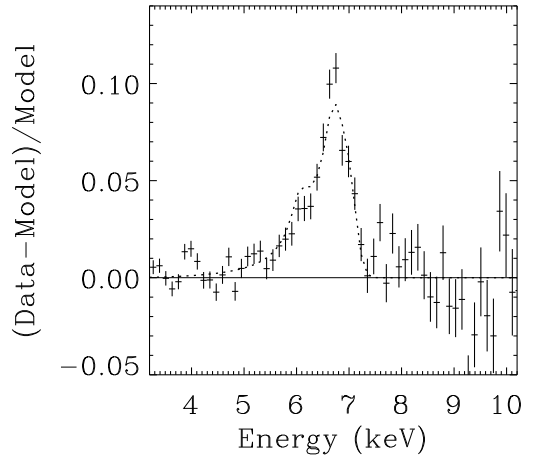}
\end{center}
\caption{Broad relativistic iron K$\alpha$ spectral line from
the inner accretion disk of a neutron star LMXB Serpens X-1
\citep{BhattacharyyaStrohmayer2007b}. The x-axis shows the X-ray energy,
and the y-axis gives the observed intensity in excess to the best-fit
continuum spectral model. The data points clearly
show a broad asymmetric spectral emission line, while the dotted profile is a
relativistic model which fits the observed line well (\S~\ref{BroadRelativisticIronLines}).
\label{diskline2_1}}
\end{figure}

Similar to black hole systems, a broad iron line from neutron star LMXBs 
has also been known for many years (see, for example, \cite{Asaietal2000}). 
However, the inner accretion disk origin of this line could not be
established until recently. This was because the characteristic asymmetry
of the relativistic nature could not be detected due to the modest
signal-to-noise ratio. In 2007,
\citet{BhattacharyyaStrohmayer2007b}, for the first time, established
the inner accretion disk origin of the broad iron line from a neutron star LMXB
(see Figure~\ref{diskline2_1}).
This was done by analyzing the {\it XMM-Newton} data from Serpens X-1.
Soon, \citet{Cackettetal2008a} confirmed this finding
using the independent {\it Suzaku} data from the same source. These authors
also reported the detection of relativistic iron lines from two other
neutron star LMXBs: 4U 1820-30 and GX 349+2. As of now, the inner disk
origin of broad iron line has been confirmed for ten neutron star LMXBs
\citep{BhattacharyyaStrohmayer2007b, Cackettetal2008a,
Pandeletal2008, DAietal2009,
Cackettetal2009a, Papittoetal2009,
Shaposhnikovetal2009, Reisetal2009,
diSalvoetal2009, Iariaetal2009,
Cackettetal2009b}.
This opens up a new way to measure neutron star parameters, as described
in \S~\ref{BroadRelativisticIronLineMethod}.

\subsubsection{Broad Relativistic Iron Line Method}\label{BroadRelativisticIronLineMethod}

The broad relativistic iron line can be used to constrain various neutron star 
parameters. This line originates from an annular region of the inner
accretion disk (see \S~\ref{WhatisaBroadRelativisticIronLine}). 
The ratio of the inner edge radius of this annulus 
($r_{\rm in, ann}$) to the neutron
star mass ($M$) can be constrained from the fitting of the iron line with a relativistic
model (see, for example, 
\cite{BhattacharyyaStrohmayer2007b}). Since the disk inner edge radius
has to be obviously greater than or equal to the neutron star radius, the measured
$r_{\rm in, ann}c^2/GM$ gives a hard upper limit on the neutron star 
$Rc^2/GM$. This can be useful to constrain the stellar $M-R$ space (see, for example,
Figure~\ref{eosrbym}).

The relativistic iron line may also be used to constrain the neutron star
angular momentum parameter, if $r_{\rm in, ann}c^2/GM$ $< 6$. Here is how.
The accretion disk can extend up to the radius ($r_{\rm ISCO}$) of the ISCO,
or the neutron star radius ($R$), whichever is greater
(see Figure 1 of \cite{Bhattacharyyaetal2000}). However, the disk
may be truncated at a larger radius by the magnetic field and/or radiative
pressure, and/or by any other means. Therefore, if $r_{\rm in}$ is the disk
inner edge radius, then $r_{\rm in, ann}c^2/GM \ge r_{\rm in}c^2/GM \ge Rc^2/GM$ and 
$r_{\rm in, ann}c^2/GM \ge r_{\rm in}c^2/GM \ge r_{\rm ISCO}c^2/GM$. 
For a non-spinning neutron star, $r_{\rm ISCO}c^2/GM = 6$, and for a spinning neutron 
star with a corotating disk, $r_{\rm ISCO}c^2/GM$ $< 6$. Therefore, if the 
best fit value of $r_{\rm in, ann}c^2/GM$ is less than 6, then that
must be because of the effect of the stellar spin on the spacetime, and
hence can, in principle, be useful to constrain the neutron star angular 
momentum parameter (see Figure 1a of \cite{Miller2007}). 
Here we note that the Kerr spacetime has been assumed
in the Figure 1a of \citet{Miller2007}, while the exterior spacetime of a spinning 
neutron star deviates from the Kerr geometry 
\citep{MillerLamb1996}.

In \S~\ref{KilohertzQuasi-periodicOscillations}, 
we have mentioned that kHz QPOs are likely associated with the inner accretion disk.
If this is true, then the relativistic iron line and the kHz QPOs are 
possibly physically related, and their simultaneous observation and joint
analysis may, not only shed light on the physics of these features, but also
be useful to constrain the neutron star parameters. For example,
\citet{Milleretal1998} suggested that the upper kHz QPO frequency ($\nu_{\rm u}$)
is close to the orbital frequency at the inner edge of the optically thick 
emission, and thus from a radius close to the $r_{\rm in, ann}$ inferred 
from the iron line. If this is true, then the neutron star mass ($M$) can be
measured from the relation $\nu_{\rm u} = (1/2\pi)\sqrt{GM/r^3_{\rm in, ann}}$
(assuming Schwarzschild spacetime; \cite{Cackettetal2008a}).

Broad relativistic iron line method has emerged as a way to constrain the
neutron star parameters only recently. Although this method has a great potential,
it suffers from systematics similar to other methods. These systematic
uncertainties may appear due to (1) the lack of knowledge of various parameters,
such as the observer's inclination angle, the nature of the hard X-ray emission
incident upon the disk, etc; and (2) the inability to model the continuum
spectrum accurately. Here we will not discuss these uncertainties in detail.
Rather, we will only note that the incident hard X-ray emission
in the neutron star LMXBs is currently believed to be originated from the
boundary layer between the inner accretion disk and the neutron star surface
\citep{Cackettetal2009b}, although more studies are required to verify this.

\subsection{Quiescent Emissions}\label{QuiescentEmissions}

\subsubsection{What is Quiescent Emission?}\label{WhatisaQuiescentEmission}

Many neutron star LMXBs have two distinctly different intensity states: (1) the
quiescent state in which the source remains for months to years; and
(2) the outburst state which continues for weeks to months (sometimes
for years; e.g., for the source KS 1731-260).
These LMXBs are called transients.
The luminosity of the source typically increases by
several orders of magnitude as it evolves from the quiescent state
into an outburst. Figure~1 of \citet{Bhattacharyyaetal2006c} 
shows the long term intensity profile of such a
transient 1A 1744-361, exhibiting three outbursts.
The transient nature of these LMXBs is normally attributed to an
accretion disk instability, which causes high accretion rates for
certain periods of time (outbursts), and almost no accretion during other times
(quiescence; see \cite{King2001} for a review). Therefore during
the quiescent state, the observed X-ray emission is expected to 
primarily originate from the neutron star surface. This emission
can be detected and measured with high sensitivity
imaging instruments on board {\it Chandra}, {\it XMM-Newton}, etc.
(e.g., \cite{Wijnandsetal2002, Cackettetal2008b, Heinkeetal2009}).

\subsubsection{Quiescent Emission Method}\label{QuiescentEmissionMethod}

As mentioned in \S~\ref{WhatisaQuiescentEmission}, the emission from
a transient neutron star LMXB in quiescence primarily comes from
the neutron star's atmosphere. This atmosphere should be composed of pure 
hydrogen, and devoid of any X-ray spectral line (see, for example, 
\cite{vanKerkwijk2004}). Moreover, the relatively low magnetic fields
of neutron stars in LMXBs (unlike isolated neutron stars, and stars in
double neutron star binaries and HMXBs) make the modeling
simpler. Therefore, assuming a blackbody emission from the neutron star surface
in a quiescent LMXB, one can constrain the stellar radius using 
equations~\ref{BBrad1} and \ref{BBrad2}. As mentioned in \S~\ref{ContinuumSpectrumMethod},
there may be effects of systematics in the estimated radius. 
During quiescence, the entire neutron star surface is expected to emit
uniformly, because the stellar magnetic field is low, and there is no
obvious effect which can create a significant asymmetry. Therefore,
The systematic uncertainty \#1 (\S~\ref{ContinuumSpectrumMethod}) may not be 
present. The source distance, which appears in equation~\ref{BBrad1},
may be known from a PRE burst from the source, or in case the source is in
a globular cluster (see uncertainty \#2; \S~\ref{ContinuumSpectrumMethod}).
The surface gravitational redshift $1+z$, which appears in
equation~\ref{BBrad2} and causes the uncertainty \#3 (\S~\ref{ContinuumSpectrumMethod}),
may be determined from an independent measurement of the neutron star
radius-to-mass ratio (see, for example, \S~\ref{SpectralLineMethod};
\S~\ref{BurstOscillationMethod}; \S~\ref{BroadRelativisticIronLineMethod}).
But, even if this ratio is not known, the neutron star $M-R$ space
can still be constrained as shown in Figure~\ref{eosr}. The color factor $f$, which
appears in equation~\ref{BBrad2} and causes the uncertainty \#4
(\S~\ref{ContinuumSpectrumMethod}), may be determined by theoretical
calculations. However, if a theoretical atmospheric spectrum for hydrogen is used for
fitting (instead of a blackbody spectrum), then a separate computation
of $f$ is not required. In fact, hydrogen atmospheric spectrum is usually
used to model the quiescent emission, which can constrain the inferred radius 
$R_\infty$ of the neutron star well (equations \ref{BBrad1} \& \ref{BBrad2}; 
\cite{Gendreetal2003, WebbBarret2007, Guillotetal2009}).

Apart from the above systematics, three observational aspects make
the quiescent emission method less reliable. (1) The interstellar 
absorption cuts off a large part of the energy spectrum for some
sources \citep{vanKerkwijk2004}. This makes the correct spectral fitting
difficult. One needs to measure the interstellar absorption independently
(for example, by simultaneous UV observation) to solve this problem.
(2) The thermal component of the energy spectrum can vary in a timescale 
of months \citep{Rutledgeetal2002}. Moreover, 
in addition to the thermal emission, sometimes a non-thermal
component is found (see, for example, \cite{Rutledgeetal2002}). This
non-thermal emission may be due to a residual accretion \citep{vanKerkwijk2004}.
Detailed observations of many
quiescent LMXBs with {\it XMM-Newton}, {\it Chandra} and future X-ray
instruments will be required to interpret the quiescent emission
X-ray spectrum correctly, and hence to increase the reliability of the method.
(3) Episodic higher accretion rates have been inferred during the 
quiescent emission \citep{Kuulkersetal2009}.
However, such high accretion is usually followed by a thermonuclear
burst \citep{Kuulkersetal2009}. Therefore if such a burst is 
detected, then the data immediately prior to it 
can be excluded from the neutron star parameter measurement analysis.

Finally we note that a neutron star heats up during accretion, and
cools down when the system goes into the quiescence. The cooling curve
of the neutron star can be useful to understand its properties (e.g., \cite{Wijnandsetal2004}).
Since repeated outbursts can be observed from a given transient system, 
the neutron star can heat up and cool down repeatedly, providing
a useful tool to study the stellar core and crust.

\subsection{Mass Measurement: Binary Orbital Motion Method}\label{MassMeasurementUsingBinaryOrbitalMotion}

The mass $M$ of a neutron star affects the motion of the companion star via gravitation,
and hence this mass may be constrained from the optical observations of the 
signatures of binary orbital motions. 
For example, the radial velocity (RV; i.e., the orbital velocity component
along the line of sight) curve of the companion, based on, say, its absorption
line spectra, can be useful to measure the binary orbital period ($P_{\rm orb}$) and the
neutron star mass function:
\begin{eqnarray}
f(M) = \frac{P_{\rm orb}K_{\rm comp}^3}{2\pi G} = \frac{M\sin^3i}{(1+q)^2}.
\label{massfunction1}
\end{eqnarray}
Here $K_{\rm comp}$ is the amplitude of the companion star RV, $i$ is the observer's
inclination angle, $q = M_{\rm comp}/M$, and $M_{\rm comp}$ is the companion mass.
Note that, since $q > 0$ and $\sin i \le 1$, $f(M)$ gives the lower limit of the neutron star
mass. Furthermore, with the estimated $i$ and $q$, $M$ can be constrained from
equation~\ref{massfunction1}. The inclination angle $i$ can be estimated from the
observed light curves, and/or from X-ray dips/eclipses that may be detected from the
LMXB (e.g., \cite{WhiteSwank1982, Parmaretal1986}). 
For example, observations of both dips and total eclipses
may imply $i \approx 75^{\rm o}-80^{\rm o}$ 
\citep{Franketal1987}. Dips and eclipses also provide the binary orbital period
(e.g., \cite{WhiteSwank1982, Parmaretal1986}). The mass ratio $q$ of
equation~\ref{massfunction1} can be 
estimated by measuring the rotational (i.e., spin-induced) broadening of the 
spectral lines originated from the companion star. This is because the spin-related
linear speed ($v_{\rm comp}$) of the companion surface can be expressed in terms of $q$
using the following formula \citep{WadeHorne1988}:
\begin{eqnarray}
v_{\rm comp} \sin i = 0.462 K_{\rm comp} q^{1/3} (1+q)^{2/3}.
\label{massfunction2}
\end{eqnarray}
However, if an estimate of $q$ is not available, then one needs to measure
the companion star mass function:
\begin{eqnarray}
f(M_{\rm comp}) = \frac{P_{\rm orb}K_{\rm NS}^3}{2\pi G} = \frac{M_{\rm comp}\sin^3i}{(1+[1/q])^2}
\label{massfunction3}
\end{eqnarray}
in order to constrain the neutron star mass. Here $K_{\rm NS}$ 
is the amplitude of the RV of the neutron-star.

Unfortunately, the optical emission of a neutron star LMXB is heavily 
dominated by the accretion disk radiation for most sources.
This optical radiation originates from the reprocessing of the X-rays
in the outer accretion disk. As a result, the
above mentioned binary orbital motion method cannot be effectively used for the 
neutron star LMXBs, except for a few sources for which the companion star can be
detected, and its relevant properties can be measured sufficiently well
(e.g., \cite{OroszKuulkers1999, Jonkeretal2005}). 
An example of such a system is Cyg X-2, 
because $\approx 70$\% of its optical emission originates from the companion star
\citep{Elebertetal2009}. The transient neutron star LMXBs in their quiescent state
might therefore be ideal to apply the above mentioned method
(see \S~\ref{QuiescentEmissions}). But, at a distance of a few kpc the 
low-mass companion stars are usually too faint to observe, again making
the binary orbital motion method ineffective.

\begin{figure}[h]
\begin{center}
\hspace*{-0.20cm}
\includegraphics[height=3.60in,angle=-90]{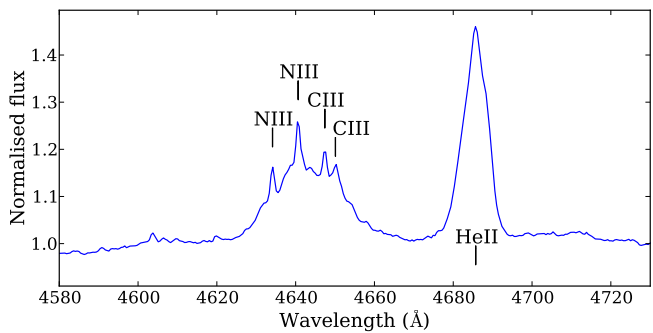}
\end{center}
\caption{Average normalised optical spectrum of Sco X-1 in the rest frame
of the companion star. The Bowen blend N III and C III emission lines from
the irradiated surface of the companion star, and the He II emission line
are shown (\S~\ref{MassMeasurementUsingBinaryOrbitalMotion}; 
figure courtesy: Danny Steeghs; \cite{SteeghsCasares2002}).
\label{Bowen}}
\end{figure}
 
\begin{figure}[h]
\begin{center}
\hspace*{-0.28cm}
\includegraphics[height=3.40in,angle=-90]{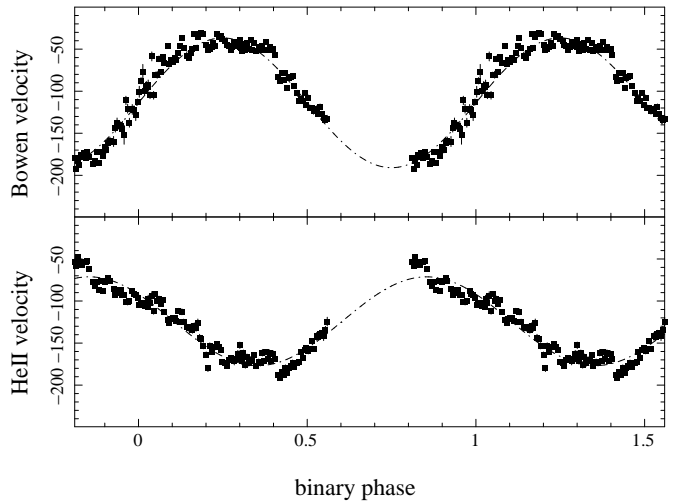}
\end{center}
\caption{{\it Upper panel}: Radial velocity (RV; in km/s) curve of the narrow 
Bowen blend emission lines (see Figure~\ref{Bowen}) from Sco X-1.
Data are repeated over two cycles, and the best sinusoidal fit is 
plotted as a dash-dot line. {\it Lower panel}: Data and sinusoidal fit of the 
RV curve of the He II emission line (see Figure~\ref{Bowen}) 
from Sco X-1. Since the Bowen blend emission lines originate
from the irradiated surface of the companion star, and the He II line
comes from near the neutron star, the RV curve of the latter is naturally
roughly in anti-phase with that of the former
(\S~\ref{MassMeasurementUsingBinaryOrbitalMotion}; 
figure courtesy: Danny Steeghs; \cite{SteeghsCasares2002}).
\label{velocities}}
\end{figure}

In 2002, \citet{SteeghsCasares2002} discovered narrow high-excitation
optical emission lines from the neutron star LMXB Sco X-1. 
The strongest lines were in the Bowen range ($4630$\AA $ - 4660$\AA), that
primarily consisted of a blend of N III and C III lines (see  Figure~\ref{Bowen}).
The N III lines originate from a UV fluorescence
process, and the C III lines are due to photo-ionization and subsequent
recombination \citep{McClintocketal1975}. These lines are believed to
originate from the irradiated surface of the companion star, mainly because
of the following reasons:
(1) the Doppler tomography of Bowen blend emission lines revealed a bright
spot of emission (see Fig. 4 of \cite{SteeghsCasares2002}); and
(2) phase-resolved spectroscopy showed that these narrow lines moved roughly in anti-phase
with the neutron-star/accretion-disk component (see Figure~\ref{velocities}).
As a result these lines led to the first estimate of $K_{\rm comp}$ for a neutron star
LMXB for which the optical emission primarily originates from the accretion disk.
Therefore, the discovery of \citet{SteeghsCasares2002} 
and the subsequent applications of the above mentioned binary orbital motion method
using the Bowen blend emission lines for several sources have 
established a new technique to constrain the neutron star masses
(see \cite{Cornelisseetal2008} and references therein;
see also \cite{MunozDariasetal2009}).

This new technique, although is very promising, suffers from systematics
mainly because of the unknown values of various source parameters. 
For example, $i$ and $q$ of equation~\ref{massfunction1} and $K_{\rm NS}$
of equation~\ref{massfunction3} are not known for many sources. 
Besides, $K_{\rm comp}$ measured from the narrow Bowen blend emission lines is only
a lower-limit to the true $K_{\rm comp}$, because these lines arise on the
irradiated side of the companion star which does not correspond to the 
center of mass of the companion. This shift of the measured $K_{\rm comp}$
from the true $K_{\rm comp}$ can be accounted for by the ``$K$-correction",
which depends on $i$, $q$ and the opening-angle of the accretion disk
(\cite{MunozDariasetal2005}; see also \cite{deJong1996}). Among these source parameters,
the uncertainty on $i$ and the opening-angle of the accretion disk
could be reduced by modeling the accretion disk, and $q$ could be
obtained by clearly resolving the narrow lines to determine the rotational broadening. 
Another systematic uncertainty comes from the poorly measured systemic
velocity, which is an important parameter to do Doppler tomography (e.g.,
\cite{Cornelisseetal2009}). Higher resolution data can be useful to reduce 
this uncertainty. Finally, the Bowen blend emission lines
may also primarily originate from the gas-stream/accretion-disk impact region,
rather than from the irradiated surface of the companion star,
for some neutron star LMXBs \citep{Elebertetal2009}.

\section{Why are Low-mass X-ray Binaries Useful?}\label{WhyareLow-massX-rayBinariesuseful}

\begin{figure}[h]
\begin{center}
\hspace*{-1.00cm}
\includegraphics[width=4.00in,angle=0]{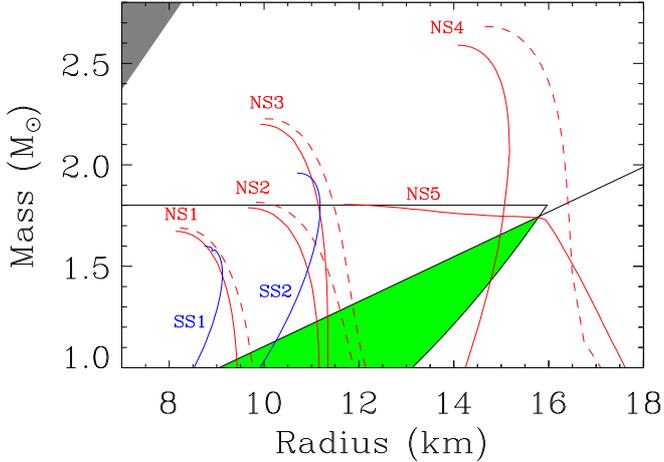}
\end{center}
\caption{This figure shows the $M-R$ space of neutron stars with the curves
corresponding to a few representative EoS models
(same as Figure~\ref{eosfirst}). 
\citet{Milleretal1999a} mentioned that the highest kHz QPO frequency
observed from the neutron star LMXB 4U 1636-536 was 1220 Hz, which
gives an allowed $M-R$ region for this source according to the 
equations \ref{freq4} and \ref{freq5} (compare this figure with Figure~\ref{eos_qpo}).
In addition, \citet{Nathetal2002} found $GM/Rc^2 < 0.163$ for 4U 1636-536 by
modeling the burst oscillations.
The green patch shows the intesection of these two allowed regions,
which should be the allowed $M-R$ space for 4U 1636-536 using the 
kilohertz quasi-periodic oscillation method 
(\S~\ref{KilohertzQuasi-periodicOscillationMethod}) and the 
burst oscillation method (\S~\ref{BurstOscillationMethod}).
Therefore, this figure demonstrates how various methods described throughout this review 
can be jointly used to tightly constrain the $M-R$ space.
However, note that the particular allowed $M-R$ space shown here may not be
very reliable, because the actual origin of kHz QPOs is not yet known, and 
the burst oscillation measurement of $M/R$ used here suffered
from systematics.
\label{eos_joint1}}
\end{figure}

As mentioned in \S~\ref{Introduction} (see also Figure~\ref{eosfirst}),
one needs to measure at least three independent parameters of the {\it same} neutron 
star in order to constrain the EoS models of the supranuclear core
matter.	Because of a number of systematics (mentioned throughout this review),
it is possible only if several measurement methods can be used for a given 
neutron star. This may be achieved for neutron star LMXB systems, because
many of them exhibit several of the following features:
thermonuclear bursts without oscillations (suitable for continuum spectrum method),
burst oscillations, PRE bursts, mHz QPOs, accretion-powered millisecond
period pulsations, kHz QPOs, broad relativistic iron lines,
quiescent emission, Bowen blend emission lines, etc. 
For example, SAX J1808.4--3658 has shown all these features except
mHz QPOs, and 4U 1636-536 has exhibited all 
these features except accretion-powered pulsations and quiescent emission.
Therefore, Bayesian method may be used to strongly constrain the 
neutron star parameter values,
which do not change from one observation to another 
(see \S~\ref{SummaryoftheMethodsUsingThermonuclearX-rayBursts}
for a discussion). This is why LMXBs are
useful systems to measure the neutron star parameters.
In Table 1 we give examples of a few sources for which various methods
mentioned in this review have been attempted, and in Figure~\ref{eos_joint1}
we demonstrate how two of these methods can be jointly used to tightly 
constrain the $M-R$ space.

There is another reason that makes neutron star LMXBs useful.
In order to understand the
nature of the supranuclear core matter extremely well, we will need to 
{\it observationally} trace out the stellar $M-R$ curve (see, for example,
\cite{OzelPsaltis2009}; see also Figure~\ref{Introduction}).
Although, currently we cannot do it, with the technological advancement
it may be doable in the future. But this will be possible only if 
neutron stars have a sufficiently large range
of mass values. Since the neutron stars in LMXBs can accrete mass
plausibly up to $0.7 M_\odot$ (e.g., \cite{vandenHeuvelBitzaraki1995}), they should
occupy a larger range in the mass space than the isolated neutron stars, and the
stars in double neutron star binaries. These latter neutron stars are formed in HMXBs, and 
hence a limited amount of matter can be accreted because of the short life of the
massive companion. In fact, 
\citet{ThorsettChakrabarty1999} found neutron star mass 
$M = 1.35\pm0.04 M_\odot$ for a sample of double neutron star binaries.
Besides, LMXBs may contain neutron stars massive enough to rule out
the softer EoS models (see \S~\ref{Introduction}).
Therefore, the LMXB systems
may be intrinsically more useful than the isolated stars, and
stars in double neutron star binary systems.

\section{Summary and Future Prospects}\label{SummaryandFutureProspects}


In this review, we have described a few methods to constrain 
the neutron star parameters. We have chosen the methods which utilize
various properties of
thermonuclear bursts, accretion-powered millisecond pulsations,
kHz QPOs, broad relativistic iron lines, quiescent emissions, and
binary orbital motions.
Some of these phenomena provide more than one method.
For example, thermonuclear burst alone provides five methods.
Throughout the review we have stressed that various systematic
uncertainties hinder the accurate measurement of the neutron star 
parameters, and a joint analysis of several phenomena
has a great potential to reduce these uncertainties. The
large amount of archival neutron star LMXB data accumulated with 
the current and past X-ray space missions can be used
for such a joint analysis, and this analysis can be further improved with
the advanced statistical and other techniques (e.g., \cite{NobleNowak2008}).
These techniques may include pulsed phase spectroscopy, higher
order timing analysis, and a higher capability spectral 
fitting recipe, which can, for example, handle the fitting
of burst continuum spectra with a large number of realistic
atmospheric models.
However, such techniques should be complemented with theoretical 
investigations of the various phenomena. For example, astrophysicists
and nuclear physicists should work more cohesively to understand
the various properties of thermonuclear bursts. Numerical study of the
inner accretion disk in order to jointly model the kHz QPOs and the
broad relativistic iron lines will also be useful. Besides,
processes of the channelled flow and the radiative transfer 
in AMPs should also be numerically investigated.

\begin{figure}[h]
\begin{center}
\includegraphics[width=3.00in]{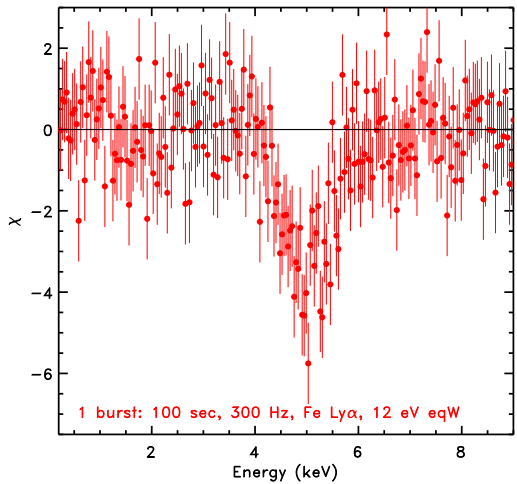}
\end{center}
\caption{The residuals ($\chi^2$ deviations) for a simulated iron Ly$\alpha$
absorption line from the neutron star surface during a thermonuclear X-ray burst.
The assumed values are the following: surface gravitational redshift $1+z
= 1.35$, line equivalent width = 12 eV, exposure = 100 s, burst blackbody
temperature = 1.7 keV and X-ray flux = $4\times10^{-8}$ ergs s$^{-1}$ cm$^{-2}$.
The simulation was done for the ``High Timing Resolution Spectrometer" (HTRS;
previously proposed for the {\it X-Ray Evolving
Universe Spectrometer} ({\it XEUS}), but now proposed for the {\it International
X-ray Observatory} ({\it IXO})). This figure shows that even a short observation
of thermonuclear X-ray bursts with HTRS may lead to a significant detection
of surface atomic spectral lines (\S~\ref{SpectralLineMethod};
figure courtesy: Didier Barret; \cite{Barretetal2008}).
\label{surface_line_sim}}
\end{figure}

\begin{figure}[h]
\begin{center}
\includegraphics[width=3.00in]{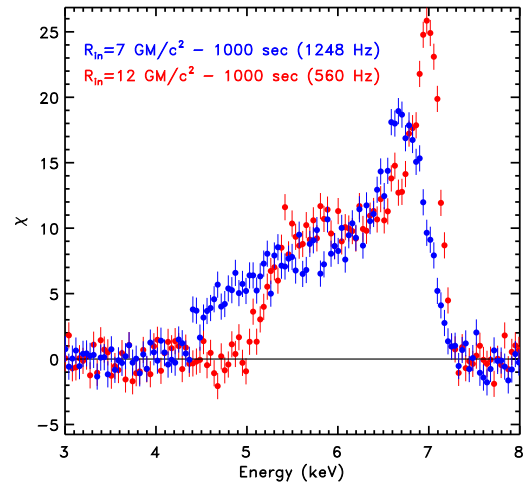}
\end{center}
\caption{The residuals ($\chi^2$ deviations) for simulated broad relativistic
iron lines at two different inner accretion disk radii. The simulation was done 
for 1000 s of HTRS exposure. This figure shows that a short HTRS observation
of relativistic lines may significantly reveal their detailed structures
(\S~\ref{BroadRelativisticIronLines}; figure courtesy: Didier Barret; \cite{Barretetal2008}).
\label{iron_emission_line_sim}}
\end{figure}

\begin{figure}[h]
\begin{center}
\includegraphics[width=3.00in]{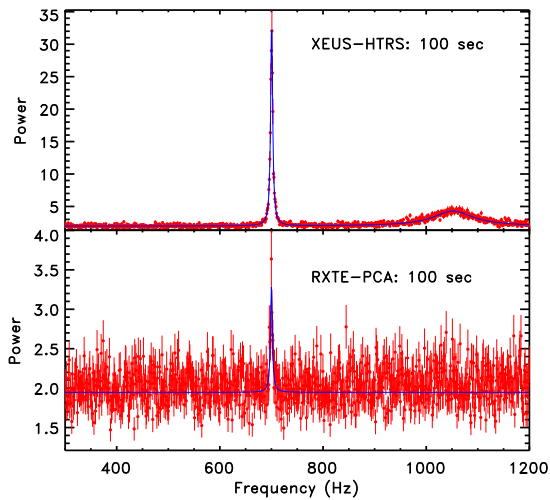}
\end{center}
\caption{A comparison of simulated power spectra observed with two instruments
(one current and one future).
The lower panel shows that the lower kHz QPO is barely detected and
the upper kHz QPO is not detected from the power spectrum created from an intensity
profile of 100 s duration observed with the {\it RXTE} PCA. 
The upper panel shows that, if the same observation were made with the HTRS,
both the kHz QPOs would be very significantly
detected (\S~\ref{KilohertzQuasi-periodicOscillations}; 
figure courtesy: Didier Barret; \cite{Barretetal2008}).
\label{khz_qpo_sim}}
\end{figure}

In addition to the archival data analysis complemented with advanced
techniques and theoretical studies, the proposed X-ray space missions 
also hold a great promise. 
The Indian multiwavelength astronomy space mission {\it Astrosat}
is planned to be launched in 2010, and 
will observe simultaneously in a wide energy range (optical to hard X-ray of
100 keV). Apart from this unprecedented capability, its LAXPC instrument
should be able to detect and measure the high frequency timing features,
such as kHz QPOs and accretion-powered pulsations, for the first time in hard X-rays
(say, up to $\approx 50$ keV). This will be useful, among other things,
to discriminate among various kHz QPO modulation mechanism models.
The Japanese X-ray satellite {\it Astro-H} is scheduled to be launched
in 2013. This mission will have an unprecedented imaging capability 
in the $0.3-80$ keV range, and will increase the reliability of the
neutron star parameter measurement methods.
The {\it International X-ray Observatory} ({\it IXO}),
jointly proposed by NASA, ESA and JAXA, will have several
highly sensitive instruments which will take the X-ray study of neutron star LMXBs
to a higher level. The launch of this space mission is planned for 2021.
The proposed ``High Timing Resolution Spectrometer" (HTRS) 
of {\it IXO} will be able to observe sources with fluxes of $10^6$ 
counts per second in the $0.3-10$ keV band without performance degradation, 
while providing good spectral resolution and excellent time resolution
(e.g., \cite{Barretetal2008}). This instrument should, therefore, be
able to reduce the systematics significantly, and measure the parameters
of several neutron stars accurately.
In Figures \ref{surface_line_sim}, \ref{iron_emission_line_sim} and \ref{khz_qpo_sim},
we show the simulated capabilities of HTRS. The {\it Advanced X-ray
Timing Array} ({\it AXTAR}), which is an X-ray observatory mission concept 
currently under study,  will provide a better sensitivity than
{\it IXO} for timing observations of accreting neutron stars (see
\cite{Rayetal2009}).
Finally, future large optical/IR telescopes, for example, the proposed 
{\it Extremely Large Telescope} \citep{Spyromilioetal2008}, 
{\it Thirty Meter Telescope} \citep{NelsonSanders2008}, can be very useful for
the binary orbital motion method, especially by detecting the companion
stars of transient neutron star LMXBs in the quiescent state.

\section*{Acknowledgments}

We thank Cole Miller for helpful suggestions which improved the manuscript,
Remon Cornelisse for useful inputs regarding the binary orbital motion method, 
Arun Thampan for providing the code used to compute the 
structure of rapidly spinning neutron stars, and Didier Barret and Danny Steeghs
for providing a few figures. We also thank the two reviewers
for their very constructive comments.
This work was supported in part by US NSF grant AST 0708424.

\clearpage

\begin{table}[h]
\centering
\begin{minipage}{184mm}
\caption{A few examples of neutron star LMXBs for which various parameter
measurement methods have been attempted. Note that this list is not at all exhaustive.}
\begin{tabular}{@{}|c|l|l|@{}}
\hline
No. & Method (Section)\footnote{Neutron star parameter measurement methods and the sections of this review in which these methods have been described.} & Source (Reference)\footnote{A few example neutron star LMXBs for which the methods have been attempted; and the corresponding example references.} \\ 
\hline
1 & Burst Continuum Spectrum & EXO 0748-676 \citep{Ozel2006}; \\
 & Method (\S~\ref{ContinuumSpectrumMethod}) & 4U 1746-37 \citep{Sztajnoetal1987}; \\
 & & 4U 1636-536 \citep{FujimotoTaam1986} \\
\hline
2 & Burst Spectral Line Method & EXO 0748-676 \citep{Cottametal2002, Ozel2006}\\
 & (\S~\ref{SpectralLineMethod}) & \\
\hline
3 & Photospheric Radius Expansion & EXO 0748-676 \citep{Ozel2006}; \\
 &  Burst Method (\S~\ref{PhotosphericRadiusExpansionBurstMethod}) & 4U 1636-536 \citep{FujimotoTaam1986}; \\
 & & X2127+119 \citep{Smale1998}; \\
 &  & Cygnus X-2 \citep{TitarchukShaposhnikov2002} \\
\hline
4 & Burst Oscillation Method (\S~\ref{BurstOscillationMethod}) & 4U 1636-536 \citep{Nathetal2002};\\
 & & XTE J1814-338 \citep{Bhattacharyyaetal2005}\\
\hline
5 & Millihertz Quasi-periodic & Not attempted yet. \\
 & Oscillation Method (\S~\ref{MillihertzQuasi-periodicOscillationMethod}) & \\
\hline
6 & Accretion-powered Millisecond & SAX J1808.4-3658 \citep{PoutanenGierlinski2003, Leahyetal2008}; \\
 & Pulsation Method (\S~\ref{Accretion-poweredMillisecondPulsationMethod}) & XTE J1814-338 \citep{Leahyetal2009} \\
 & & \\
\hline
7 & Kilohertz Quasi-periodic & 4U 1636-536 \citep{Milleretal1999a}; \\
 & Oscillation Method (\S~\ref{KilohertzQuasi-periodicOscillationMethod}) & Sco X-1 \citep{Abramowiczetal2003, StellaVietri1999}; \\
 & & GX 5-1 \citep{StellaVietri1998}; \\
 & & GX 17+2 \citep{StellaVietri1998}; \\
 & & SAX J1808.4-3658 \citep{Zhang2009} \\
\hline
8 & Broad Relativistic Iron Line & Serpens X-1 \citep{BhattacharyyaStrohmayer2007b}; \\
 & Method (\S~\ref{BroadRelativisticIronLineMethod}) & 4U 1820-30 \citep{Cackettetal2008a}; \\
 & & GX 349+2 \citep{Cackettetal2008a}; \\
 & & Cygnus X-2 \citep{Cackettetal2009b} \\
\hline
9 & Quiescent Emission Method & Sources in the globular clusters: omega Cen, M13, NGC 2808, NGC 6304\\
 & (\S~\ref{QuiescentEmissionMethod}) & \citep{WebbBarret2007, Guillotetal2009} \\
\hline
10 & Binary Orbital Motion Method & EXO 0748-676 \citep{MunozDariasetal2009}; \\
 & (\S~\ref{MassMeasurementUsingBinaryOrbitalMotion}) & 4U 1822-37 \citep{MunozDariasetal2005}; \\
 & & 4U 1254-69 \citep{Barnesetal2007} \\
\hline
\end{tabular}
\end{minipage}
\end{table}


\end{document}